\PassOptionsToPackage{sort&compress}{natbib}

\documentclass[12pt,A4paper,fleqn]{elsarticle}
\usepackage{graphicx}
\usepackage{amsmath}
\usepackage[utf8]{inputenc}
\usepackage{hyperref}
\usepackage[sort&compress]{natbib}
\usepackage{subfig}
\graphicspath{ {images/} }
\usepackage[acronym]{glossaries}
\usepackage[Export]{adjustbox}
\makeglossaries
\usepackage{bibunits}
\usepackage{booktabs}
\usepackage[acronym]{glossaries}
\usepackage{epstopdf}

\journal{Mechanical Systems and Signal Processing}
\begin{document}
\begin{frontmatter}


\title{Dynamics of space debris removal: A review}



\author[inst1,inst2]{Mohammad Bigdeli}
\author[inst1,inst2]{Rajat Srivastava}
\author[inst1,inst2]{Michele Scaraggi}

\affiliation[inst1]{organization={Department of Engineering for Innovation, University of Salento}, 
            city={Lecce},
            postcode={73100 },
            country={Italy}}

\affiliation[inst2]{organization={Center for Biomolecular Nanotechnologies, Istituto Italiano di Tecnologia - IIT}, 
            city={Arnesano},
            postcode={73010}, 
            country={Italy}}

\large\begin{abstract}
\newacronym{gps}{GPS}{Global Positioning Systems}
\large	Space debris, also known as "space junk," presents a significant challenge for all space exploration activities, including those involving human-onboard spacecraft such as SpaceX's Crew Dragon and the International Space Station. The amount of debris in space is rapidly increasing and poses a significant environmental concern. Various studies and research have been conducted on space debris capture mechanisms, including contact and contact-less capturing methods, in Earth's orbits. While advancements in technology, such as telecommunications, weather forecasting, high-speed internet, and \acrshort{gps}, have benefited society, their improper and unplanned usage has led to the creation of debris. The growing amount of debris poses a threat of collision with the International Space Station, shuttle, and high-value satellites, and is present in different parts of Earth's orbit, varying in size, shape, speed, and mass. As a result, capturing and removing space debris is a challenging task. This review article provides an overview of space debris statistics and specifications, and focuses on ongoing mitigation strategies, preventive measures, and statutory guidelines for removing and preventing debris creation, emphasizing the serious issue of space debris damage to space agencies and relevant companies.

\end{abstract}

\begin{graphicalabstract}
\end{graphicalabstract}

\begin{highlights}
    \item This review examines the current state of debris in Earth's orbit.
    \item Design parameters for the development of efficient debris capturing systems are explored.
    \item Mitigation strategies for debris removal are presented.
    \item International guidelines and preventive measures to minimize debris creation are discussed.
\end{highlights}



\end{frontmatter}



\newacronym{geo}{GEO}{Geostationary Earth Orbit}
\newacronym{leo}{LEO}{Low Earth Orbit}

\section{\large	Introduction}
\label{sec:sample1}
\large	In October 1957, an artificial satellite “Sputnik 1” was launched. The unanticipated success ignited the curiosity towards space exploration. However, due to complexity, only a few countries have capability of launching satellites and operating them. Nevertheless, there is a rapid increase in space traffic because of governmental, commercial, and military activities(telecommunication, broadcasting, weather forecasting, global positioning systems, spying satellites etc.) which leads to a gradual increase in space junk, or debris \cite{kuznetsov2015yakov,castronuovo2011active,bombardelli2011ion,dolado2015review}. Space debris is typically made up of defunct object such as abandoned satellites, fragments of rockets and spacecraft, and countless broken machine parts, all of which pose a serious threat to human space activities. As a result these threats, the International Space Station, the space shuttle, and many satellites often need to perform orbital maneuvers to avoid collisions with debris \cite{fukushige2007comparison,schaub2014active,liou2015usa,schaub2015cost,rossi2005earth}. There is no doubt that, with the advancement of space technology, the domain of human activities in space is increasing. However, many problems are also arising due of it. For example, the increasing space activities also leads to the increase in space debris, which poses concern for environmental pollution. 
\cite{adilov2015economic,adilov2020economics,adushkin2020small,jakhu2007legal,adilov2013earth}.

\begin{figure}[h]
    \centering
    \includegraphics[width=0.9\linewidth]{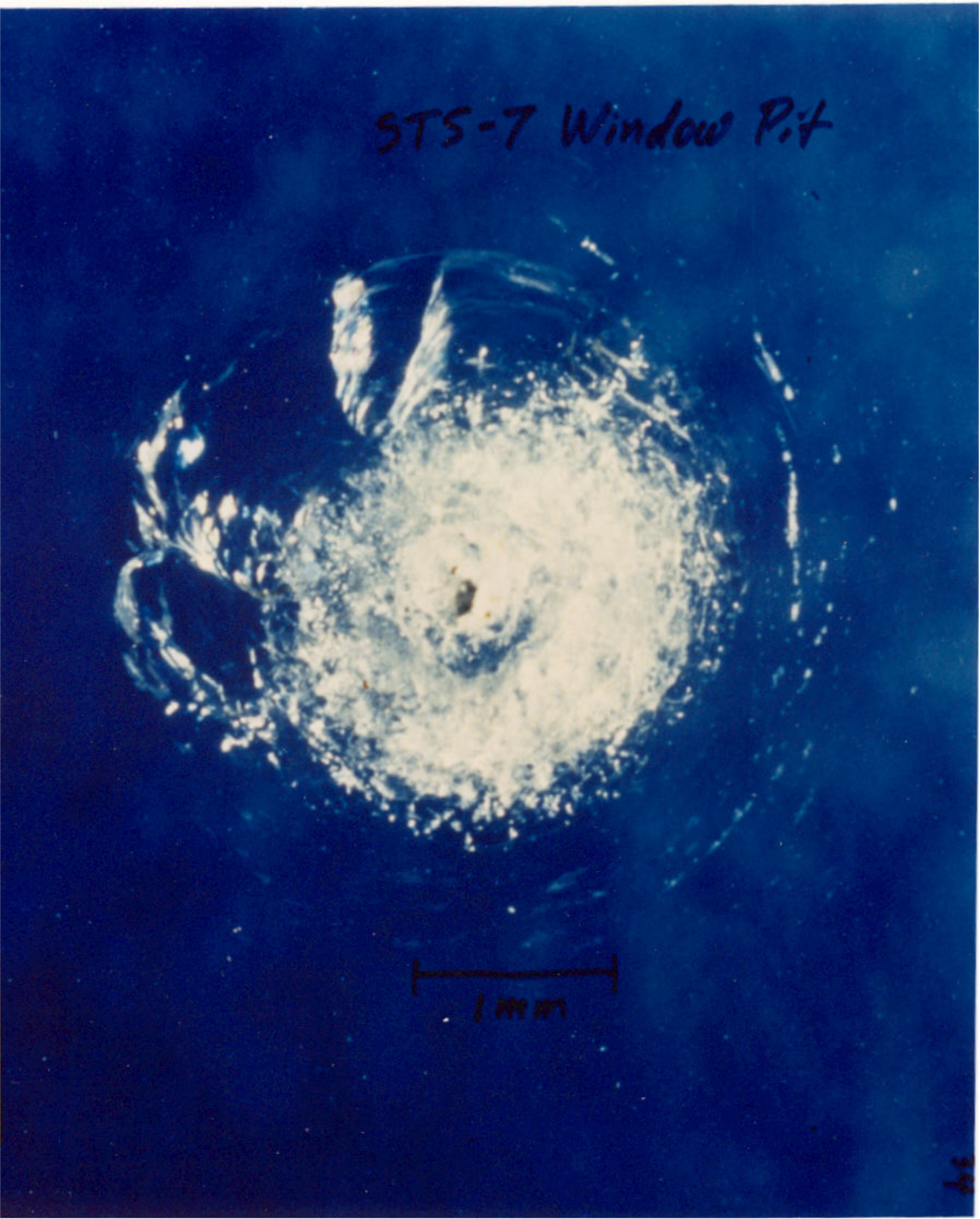}
    \caption{Damage caused by a 200-micron paint flake to the window of STS-7. Credit: NASA \cite{nasaaphoto}.}
    \label{fig:window_pit}
\end{figure}

Since the beginning of the space age in 1957, we have launched approximately 6200 rockets into Earth's orbit, with around 5700 operational satellites out of a total of 8410 deployed satellites \cite{spaceenvironmentstatistics}. These satellites are placed in various Earth's orbits, including low Earth orbit (\acrshort{leo}), and geostationary Earth orbit (\acrshort{geo}). The area located in space below an altitude of 2,000 km from the Earth's surface is considered low Earth orbit (\acrshort{leo}). It is the closest orbit and contains the most space debris compared to other Earth's orbits. The close proximity of LEO is advantageous for satellites to capture high resolution images of Earth surface. This orbit can often be used to track the location of objects for global positioning systems as well as military applications. The International Space Station is also located in this orbit to make the travel of astronauts easier. The geostationary Earth orbit (\acrshort{geo}) is located at an altitude of 36,000 km from the Earth's surface, \cite{johnson1984history,hampf2013ground} and the orbital period of the objects in the GEO orbit is equal to the Earth's rotation period. Thus, the speed of the objects in the Earth's orbits is related to the orbital period. For example, the speed of a satellite or any other object, including debris in LEO and GEO, is around 7.8 km/s and 3.1 km/s, respectively \cite{esa_1111}.  The high speed of space debris in orbit is a major concern because it poses a risk of colliding with spacecraft or satellites, potentially causing significant damage \cite{tran2022developing,WinNT,gordon2021international}. For instance, during the Space Transportation System (STS) program, a 4 mm pit was found on the STS-7 window (see Fig. \ref{fig:window_pit}), which was created by the impact of a paint flake with a diameter of approximately 200 microns\cite{hall2014history, reynolds1989orbital}. During the entire STS program, the fleet flew more than 60 missions and experienced 177 impacts on its outer windows, 45 of which required window replacement \cite{reynolds1989orbital}.

\begin{figure}[h]
    \centering
\includegraphics{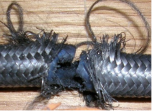}
    \caption{The power cable of ISS, damaged by the collision of debris \cite{spacecraft-damage_2011}.}
    \label{fig:power_cable}
\end{figure}

Studies also suggested that even aluminum oxide (Al$_2$O$_3$) released from the exhaust of a rocket can form micrometer-sized dust and mm-to-cm-sized slag particles. These tiny debris (dust or slag particles) has a high risk of collision with satellites or other objects in orbit \cite{garcia_2015,celestino2004orbital,esa-esa,national-aeronautics}. It is often reported that these tiny, fast objects have damaged various systems in space \cite{garcia_2015,esaaaa,christiansen2004space}. A similar damage is reported in the Fig. \ref{fig:power_cable} where the debris severely damaged the power cable on the International Space Station (ISS) \cite{spacecraft-damage_2011}.

\begin{figure}[h]
    \centering
    \includegraphics[width=0.9\linewidth]{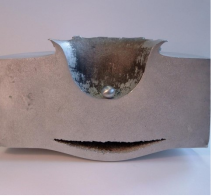}
    \caption{Damage caused by the collision of a small sphere of aluminum with a block of aluminum. Credit: ESA photo\cite{esa-5}.}
     \label{fig:collision_Al}
\end{figure}

\begin{figure}[h]
    \centering
    \includegraphics[width=0.9\linewidth]{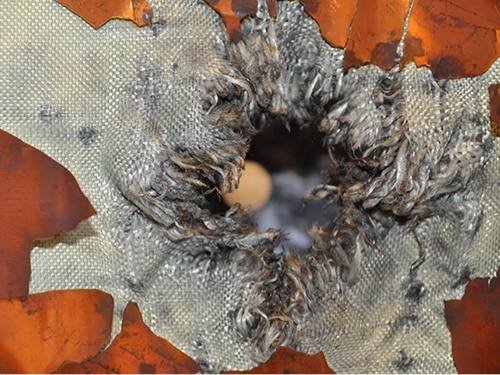}
    \caption{Test showing the damage caused by the collision of a bullet with ATV's Kevlar-Nextel fabric. Credit: ESA photo \cite{peter-b}.}
     \label{fig:bullet_ATV}
\end{figure}

\begin{figure}[h]
    \centering
    \includegraphics[width=0.9\linewidth]{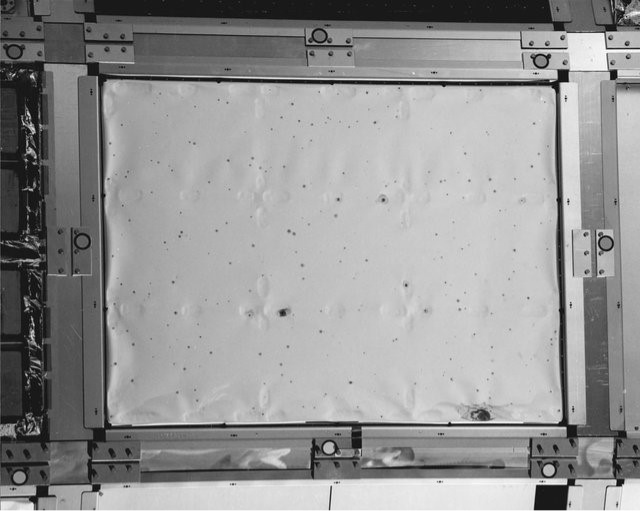}
    \caption{The impact of tiny objects colliding with a silver Teflon™ blanket on "LDEF". Credit: NASA JSC \cite{nasaaphoto}.}
     \label{fig:Silver_Teflon}
\end{figure}

\begin{figure}[h]
    \centering
    \includegraphics[width=0.9\linewidth]{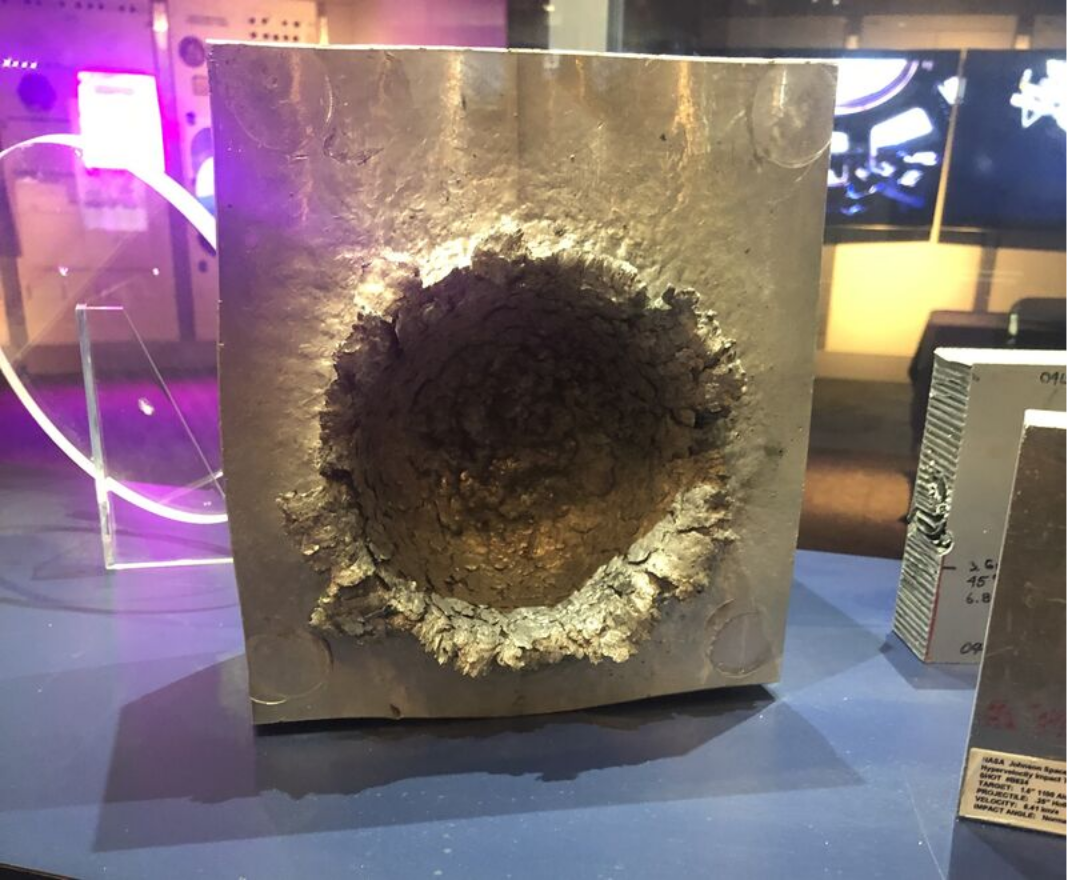}
    \caption{A crater is created on a 4-inch-thick aluminum block by the collision of a 1-inch, half-ounce plastic cylinder in orbit. Credit: NASA\cite{nasa_2015}.}
     \label{fig:plastic_Aluminium}
\end{figure}

At the beginning of the space program, the general view was that space was large enough to accommodate everything without getting filled up. However, this perception was wrong because space debris is growing at an enormous rate, and the fast-moving debris in the orbits collides with other objects, leading to the creation of more debris and thus filling the orbital space at an alarming rate.\cite{aslanov2019gravitational,oda2014optical,zhang2019high}.
\newacronym{isa}{ISA}{International Space Agency}
\newacronym{atv}{ATV}{Automated Transfer Vehicle}
\newacronym{gpa}{GPa}{Giga Pascal}
To understand the impact of fast-moving debris on the orbiting satellites and spacecraft, the International Space Agency(\acrshort{isa}) conducted several tests. In such a test, they collided a small sphere of aluminum with a diameter and mass of 1.2 cm and 1.7 g, respectively, traveling at a speed of approximately 6.8 m/s into an 18 cm-thick block of aluminum. The study reported that a crater with a diameter and depth of 9.0 cm and 5.3 cm, respectively, was created on the aluminum block (see Fig.\ref{fig:collision_Al}) due to the impact of a small aluminum sphere. This impact also generated a pressure of around 365 GPa and raised the temperature of the aluminum block to 6000K\cite{tarran2021prepare,esa-5}. In another ISA test (shown in Fig. \ref{fig:bullet_ATV}), an aluminum bullet (7.5 mm) traveling at 7 m/s was fired at a bullet-proof-vest type fabric. This fabric resembles the outer skin of an automated transfer vehicle (\acrshort{atv}) and is used to analyze the collision impact of small debris with the spacecraft \cite{peter-b,pedrick1976using}.
The likelihood of debris colliding with spacecraft and satellites is determined by the amount of time they spend in orbit and their size. Therefore, larger spacecraft that spend more time in orbit are more vulnerable. The vulnerability of spacecraft has been studied by placing the "Long Duration Exposure Facility (LDEF)" in low Earth orbit for 69 months. Fig. \ref{fig:Silver_Teflon} illustrates the damage caused by the impact of tiny debris on a silver Teflon$^{\text{TM}}$ blanket on "LDEF" \cite{belk1997meteoroids,national1995orbital}. Similarly, severe damage due to the impact of a 1-inch, half-ounce plastic cylinder at a velocity of about 24462 km/h with a 4-inch-thick aluminum block is reported in a study \cite{nasa_2015} and shown in Fig. \ref{fig:plastic_Aluminium}.

\newacronym{iss}{ISS}{International Space Station}
As previously stated, the presence of debris in orbit can have a significant impact on human space activities. Therefore, keeping track of these objects or debris is essential. There are many methods for tracking these objects or debris. The space surveillance network (SSN), which includes radars and optical telescopes, can easily track larger debris or objects larger than 10 cm in size \cite{garcia_2015}. The medium-sized debris (between 1 cm and 10 cm) is often traceable by today's advanced technology. However, the detection of small-sized debris (between 1 mm and 1 cm) is still a challenging task. The estimation of small-size debris is done by analyzing  and statistically evaluating the damages on the return spacecraft, or, as mentioned earlier, with the use of long-duration exposure facility (LDEF). 
\newacronym{ssn}{SSN}{Space Surveillance Network}
\newacronym{esa}{ESA}{European Space Agency}

As per the Space Debris Environment Report issued by \acrshort{esa} (updated on May 10, 2022), the total mass of all space objects is more than 9900 tonnes \cite{spaceenvironmentstatistics}. 
 The report also highlighted that the population of large-sized debris is around 34000; that of medium-sized debris is around 90,000; and the statistically estimated population of small-size debris is around 128 million.
\newacronym{u.s.}{U.S.}{United States}

\newacronym{ssn}{SSN}{Space Surveillance Network}

\begin{figure}[h]
    \centering
    \includegraphics[width=0.9\linewidth]{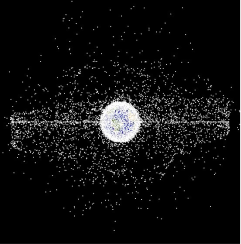}
    \caption{The populations of debris in LEO and GEO orbits. Credit: NASA ODPO\cite{nasaaa}.}
    \label{fig:pop_Debris_LEO_GEO}
\end{figure}

More than 27,000 objects are regularly tracked by \acrshort{ssn}, and their number is increasing as these objects collide with each other or with spacecraft (Fig.\ref{fig:pop_Debris_LEO_GEO} presents the population of space debris in LEO and GEO) \cite{nasaaa}. During the last decade, the amount of debris in orbit has increased dramatically (almost doubled). Two major events that contributed to this upward trend are: (1) In 2007, China conducted an anti-satellite missile test on the satellite Fengyun-1C, resulting in the largest debris cloud ever produced by a single event (increasing the debris population by nearly 25\%). (2) In 2009, a collision between Iridium-33 and Cosmos-2251, a defunct Russian satellite, produced 2296 traceable debris and hundreds of thousands of untraceable pieces of junk, posing a threat to other satellites orbiting nearby \cite{laporte2008operational,payne1997first,kamm2015secure,arzelier2020rigorous,jakhu2010iridium,mejia2009collision,hildreth2014threats,kelso2009space}.
According to Kessler and Cour-Palais (1978), if the orbital debris keeps on increasing, then a time will come when the debris population might reach a critical density above which cascading collisions occur, which could self-sustain even without future launches, rendering the space environment useless for hundreds to thousands of years. Many researchers discussed this issue, and it was quoted as "Kesseler Syndrome".

\begin{figure}[h]
    \centering
    \includegraphics[width=0.9\linewidth]{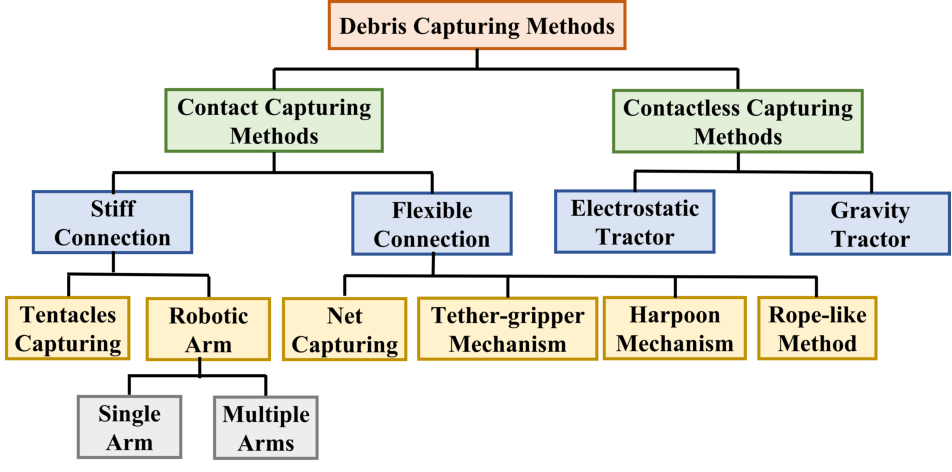}
    \caption{Methods of capturing space debris. "Adapted from Progress in Aerospace Sciences , Vol 80, Minghe Shan and Jian Guo and Eberhard Gill, Review and comparison of active space debris capturing and removal methods, 18-32., Copyright (2016), with permission from Elsevier" \cite{shan2016review}.}
    \label{fig:methods_of_debris_capture}
\end{figure}

\begin{figure}[h]
    \centering
    \includegraphics[width=0.9\linewidth]{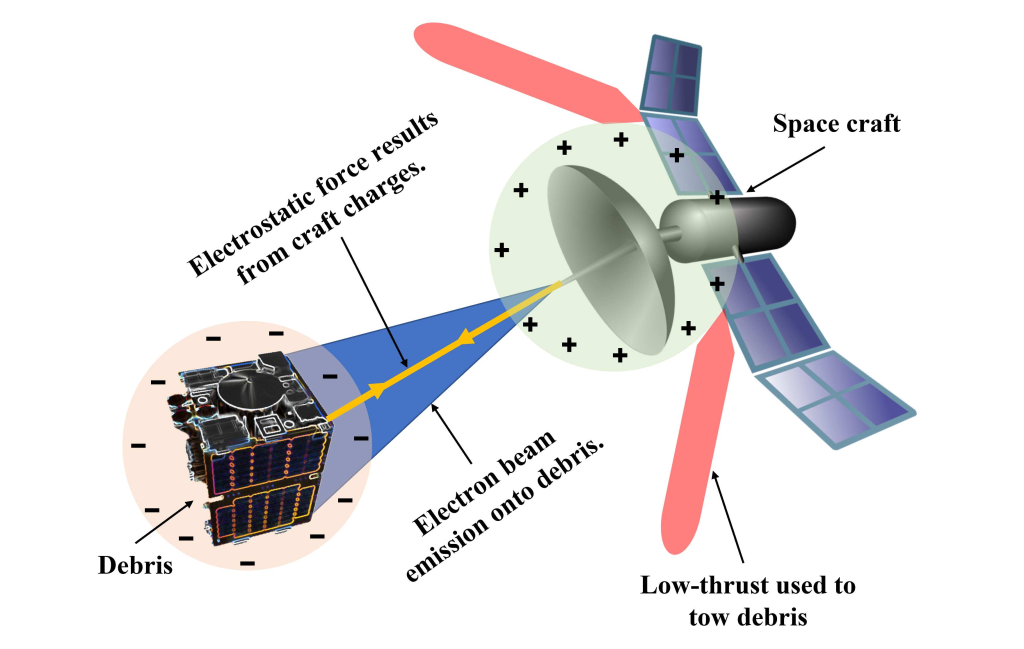}
    \caption{Schematic of a contact-less method for the removal of space debris.
    "Adapted from Advances in Space Research, Vol 62, Trevor Bennett and Hanspeter Schaub, Contactless electrostatic detumbling of axi-symmetric GEO objects with nominal pushing or pulling, 2977-2987, Copyright (2018), with permission from Elsevier"
\cite{BENNETT20182977}).}
    \label{fig:contact_less_removal}
\end{figure}

In recent years, several methods have been suggested in order to capture and remove space debris \cite{benvenuto2015dynamics,nishida2011strategy,botta2016evaluation,shan2021post,reed2013development,ribeiro2018evolution,guang2012space,mayorova2021analysis,liu2021analysis,nishida2018lightweight}. These methods are divided into two main categories: contact capturing methods and contact-less capturing methods\cite{ribeiro2018evolution,shan2016review}. The framework of methods for capturing space debris is depicted in Fig. \ref{fig:methods_of_debris_capture}.
In the contact-less debris removal method, an external force has been applied in a controlled manner to de-orbit debris, resulting in its atmospheric re-entry (see Schematic \ref{fig:contact_less_removal}). For example, an energetic plasma beam has been projected from a spacecraft towards the debris. This creates an opposite charge on the debris, resulting in its deceleration, hence facilitating the re-entry of space debris into the Earth's atmosphere, where it would burn up \cite{saini2022numerical,takahashi2018demonstrating,phipps2012removing}.
Other contact-less capture methods include the gravity tractor \cite{shan2016review,lu2005gravitational} or an electrostatic tractor \cite{shan2016review,murdoch2008electrostatic}, which use external forces such as gravitational or electrostatic forces to deflect the space debris.
Lu and Love \cite{lu2005gravitational} presented a conceptual design of a spacecraft that uses gravitational force to alter the trajectory of an asteroid. In their design, they proposed that the spacecraft hover near the asteroid with its thrusters directed outwards, deflecting the asteroid from its trajectory. This method is unaffected by the structure, surface, or rotation state of the asteroid \cite{lu2005gravitational,bombardelli_pelaez_2011}. The electrostatic tractor approach uses Coulomb forces to de-orbit space debris. In this method, a spacecraft (a tug vehicle) approaches space debris (a rocket body or a dead satellite) and emits an electron beam at it. The target (debris) gets negatively charged due to the storage of additional electrons, while the tug vehicle that emits electrons becomes positively charged. Thus, the resultant attractive force between two opposite charged bodies (the tug vehicle and the debris) is used to initiate the de-orbiting process of debris without any physical contact \cite{bengtson2018survey,schaub2012geosynchronous,schaub_sternovsky_2014}.

\begin{figure}[h]
    \centering
    \includegraphics[width=0.7\linewidth]{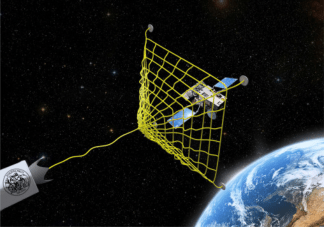}
    \caption{An illustration of a contact capturing method for the removal of space debris using a tethered-net. "Reprinted from Acta Astronautica, Vol 110, Riccardo Benvenuto and Samuele Salvi and Michèle Lavagna, Dynamics analysis and GNC design of flexible systems for space debris active removal, 247-265, Copyright (2015), with permission from Elsevier"
    \cite{benvenuto2015dynamics}.}
    \label{fig:contact_capture_illustration}
\end{figure}

On the other hand, contact capturing methods for the removal of space debris (see illustration \ref{fig:contact_capture_illustration}) are classified into two main, the stiff and the flexible connection methods \cite{benvenuto2015dynamics},\cite{guang2012space},\cite{fehse2014rendezvous},\cite{blackerby2019elsa}. 
Tentacle and robotic arm capture methods are the two broad categories of stiff connection capture method. The tentacle capture method features flexible appendages that grasp the debris at several points rather than latching on to a single point, as in the case of the robotic arm capturing method. The tentacle capture method can be used with or without the utilization of a robotic arm. In the case of the robotic arm tentacle capture method, a clamping mechanism is used to capture the object or debris. Then it performs a controlled re-entry of debris into the atmosphere, where the debris is finally burned up\cite{biesbroek2012deorbit}. On the contrary, the tentacle capturing without a robotic arm is based on a docking mechanism. The docking clamp on the chaser is designed according to the shape of the targeted debris\cite{shan2016review}. Thus, debris is locked within the chaser's clamp and is ultimately de-orbited. The tentacle capturing methods are easy to test and have a higher Technology Readiness Level (TRL). However, these methods are not cost-effective \cite{shan2016review}. 

\begin{figure}[h]
    \centering
    \includegraphics[width=0.9\linewidth]{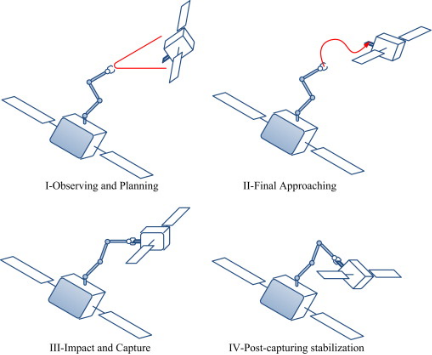}
    \caption{Steps showing the capture mechanism of a robotic arm. "Reprinted from Progress in Aerospace Sciences, Vol 68, Angel Flores-Abad and Ou Ma and Khanh Pham and Steve Ulrich, A review of space robotics technologies for on-orbit servicing, 1-26, Copyright (2014), with permission from Elsevier"
    \cite{flores2014review}.}
    \label{fig:Robotic arm}
\end{figure}

The robotic arm capture method is used for custom-designed satellites that have installed markers. These markers can be easily gripped by the end effector of the robotic arm. Thus, this mechanism is highly reliable and reusable \cite{HAN2021}. The capturing mechanism of the robotic arm works using the following steps (illustrated in Fig. \ref{fig:Robotic arm})\cite{flores2014review}:
Firstly, a camera captures images of the debris at different angles. Then, the obtained information has been processed to  analyze the targeted debris's physical features, orientation, and position. Once the physical information is obtained, the installed marker is located and the capturing phase initiates. The robotic arm seizes the target from the marker and stabilizes it using the de-tumbling mechanism\cite{flores2014review,WEN20151451}. 

\begin{figure}[h]
    \centering
    \includegraphics[width=0.9\linewidth]{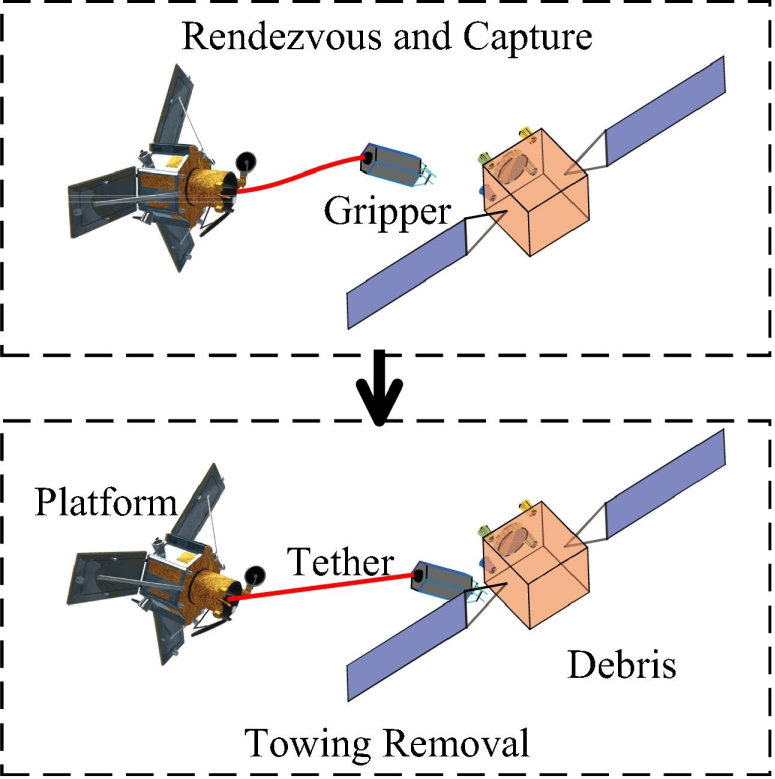}
    \caption{A schematic showing a tethered gripper debris capture and removal mechanism.
    "Reprinted from Advances in Space Research, Vol 64, Issue 6, Cheng Jia and Zhongjie Meng and Panfeng Huang, Attitude control for tethered towing debris under actuators and dynamics uncertainty, 1286-1297, Copyright (2019), with permission from Elsevier"
    \cite{jia2019attitude}.}
    \label{fig:tethered_gripper}
\end{figure}

In case of flexible connection capturing methods such as tether-nets and tether-grippers, objects are captured without a locating a specific point. The tethered-net capturing method is one of the most promising ones because of its various advantages. These include the ability to capture debris in a variety of shapes and dimensions; the ability to even capture debris when the chaser satellite and the target are at a large distance; and the ability to do so in a cost-effective manner.
The capturing mechanism of tether-net involves the following steps: net deployment, target capturing, and de-orbiting \cite{shan2019contact,trivailo2016dynamics,shan2017deployment}. In the net deployment stage, the flying weights are attached to the corners of the net, and these weights are shot up by a spring system called the "net gun". During the deployment of the net, these flying weights expand the net and wraps around the targeted debris. Finally, the captured debris is deorbited \cite{shan2017deployment}.
Another flexible connection capturing method is the tethered harpoon method, in which the chaser spacecraft fires a harpoon at a non-operational satellite or debris from a suitable distance. The harpoon penetrates and anchors the debris. Then, the chaser spacecraft tows the debris into the atmosphere, where burns it up \cite{dudziak2015harpoon}. This method has a high risk of creating a new fragments because of the harpoon penetration into the debris \cite{billot2014deorbit,forshaw2017removedebris,reed2012grappling,shan2016review}.
The tether-gripper mechanism is also a flexible connection-capturing method similar to net capturing. In this method, a spacecraft maneuvers near the debris to determine the best possible capturing position.  Once the position is determined, the gripper is moved towards the planned position and physically grabs the debris. Then, the debris is transported into the disposal orbit by the thrust of the spacecraft (see schematic \ref{fig:tethered_gripper}) \cite{bonnal2013active,jia2019attitude}.

\begin{figure}[h]
    \centering
    \includegraphics[width=0.9\linewidth]{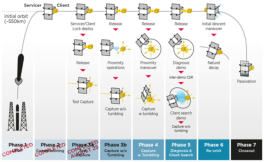}
    \caption{Schematic demonstrating testing phases of ELSA-d \cite{astroscale_2021}.}
    \label{fig:Elsa_d}
\end{figure}

Some space companies, such as Astroscale, are working on developing innovative solutions for the removal of space debris \cite{blackerby2019elsa,forshaw2020active}. On August 25th, 2021, in an initial phase of testing, Astroscale successfully demonstrated the capability of ELSA-d (End-of-Life Services by Astroscale-demonstration) to capture its client satellite using the docking-based magnetic capture method \cite{astroscale_2021}. The testing phases of Astroscale's ELSA-d are presented in the schematic \ref{fig:Elsa_d}.
The European Space Agency (ESA) also plans to launch a mission to capture and remove space debris in 2025. The ESA's ClearSpace-1 mission is jointly built by ClearSpace, a Swiss startup, and a team of researchers from the Ecole Polytechnique Fédérale de Lausanne (EPFL). The purpose of the ClearSpace-1 mission is to use a tentacle capture mechanism to capture Vespa (Vega Secondary Payload Adapter). This payload adapter was left behind by ESA's Vega launcher in 2013, with a weight of around 100 kilograms and located about 800 kilometers above Earth \cite{esa-1}.  Apart from these developments by various space agencies, the actual removal of debris has not yet taken place. The slow process of development of space cleaners is a serious concern because the increasing population of debris possibly creates insufficiency in orbital space for future launches \cite{lemmens2020esa,shan2016review,busche2020controllable, lewis2009active}.

The purpose of this study is to clarify the critical state of space debris in orbit and provide insight towards the design and development of an efficient debris capture and removal process. The article is organized into the following sections: In Section 2, we discuss the present situation of debris in Earth's orbit. The important parameters for the design of an effective capture and removal system are presented in Section 3. Further, various methods and models for the  removal of debris, as mitigation strategies, are discussed  in Section 4. In addition Section 5, discusses about the preventive measures to impede the growth of debris in orbit. Finally, an overall conclusion is drawn.

\section{\large Current status of Debris in Earth's orbit}
\begin{figure}[h]
    \centering
    \includegraphics[width=0.9\linewidth]{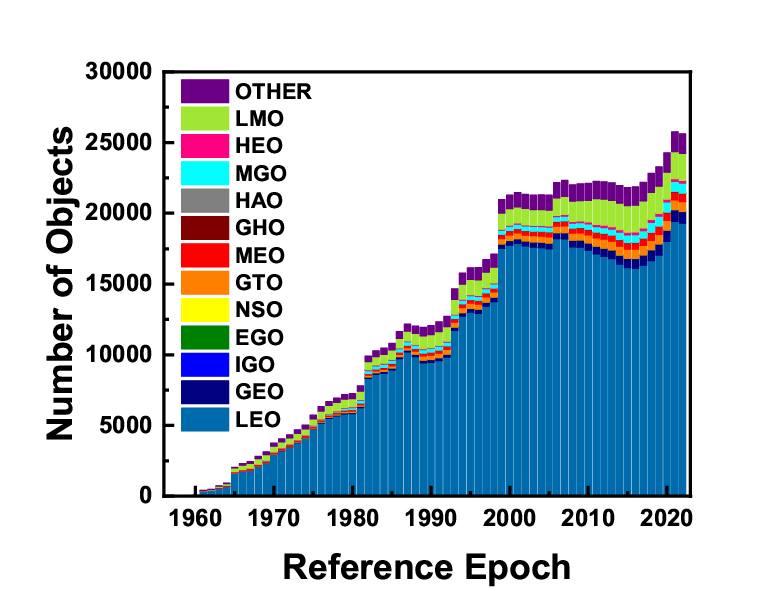}
    \caption{Evolution of number of objects in orbit \cite{esa_report2022}.}
    \label{fig:Evolution of number}
\end{figure}

\begin{figure}[h]
    \centering
    \includegraphics[width=0.9\linewidth]{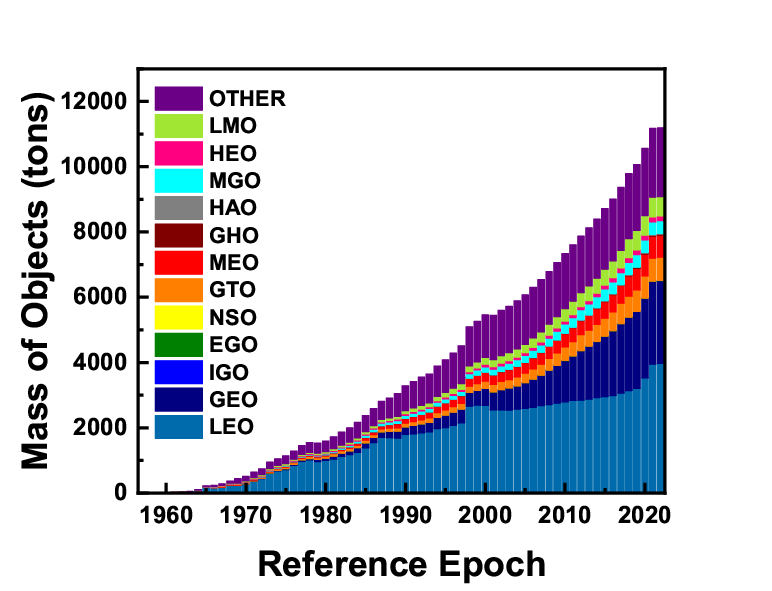}
    \caption{Evolution of mass of objects in the Earth's orbit\cite{esa_report2022}.}
    \label{fig:evolution of mass}
\end{figure}

\begin{figure}[h]
    \centering
    \includegraphics[width=0.9\linewidth]{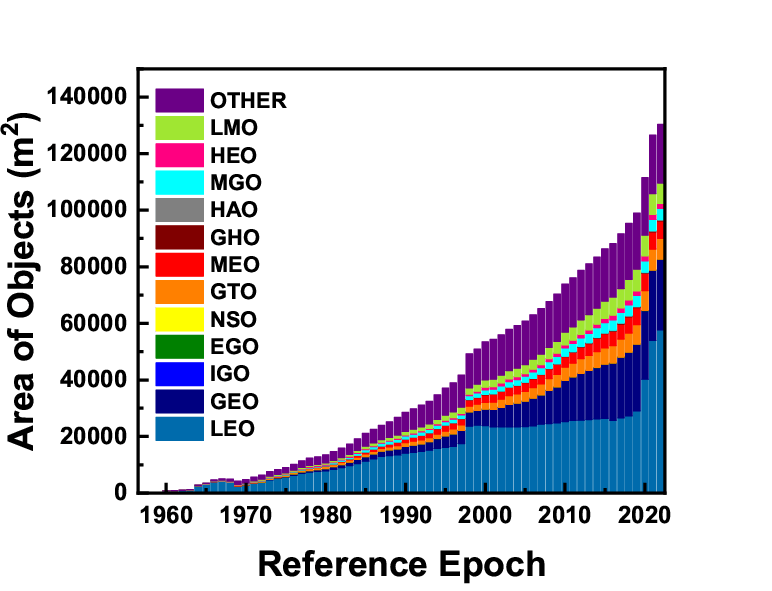}
    \caption{Evolution of area in the Earth's orbit reported by ESA\cite{esa_report2022}.}
    \label{fig:area_ESA}
\end{figure}

In this section, we discuss the current status of recorded debris in the Earth's orbit as per the database presented by the various space agencies, including \acrshort{esa}'s Space Debris Office at Darmstadt, Germany and  United State Space Surveillance Network. Historical evidence presented by the \acrshort{esa} shows that the number of objects (data limited to the capability of the surveillance system of the epoch) in Earth's orbit is increasing continuously (see Fig. \ref{fig:Evolution of number}), especially in low Earth orbit\cite{esa_report2022}. 
The increasing number of objects leads to a steady increase in the total mass of the debris. Fig. \ref{fig:evolution of mass} reports the increasing trends of mass in the Earth's orbits over the years \cite{esa_report2022}. As seen from the figure \ref{fig:evolution of mass}, the total mass of objects is around 10,000 tonnes and is mainly distributed in the LEO and GEO (more than 6000 tonnes). 
The area occupied by these objects has also increased exponentially over the years and covers almost 120,000 m$^2$ (year 2022) of the Earth's orbits (see Fig. \ref{fig:area_ESA}). 

\begin{figure}[h]
    \centering
    \includegraphics[width=0.9\linewidth]{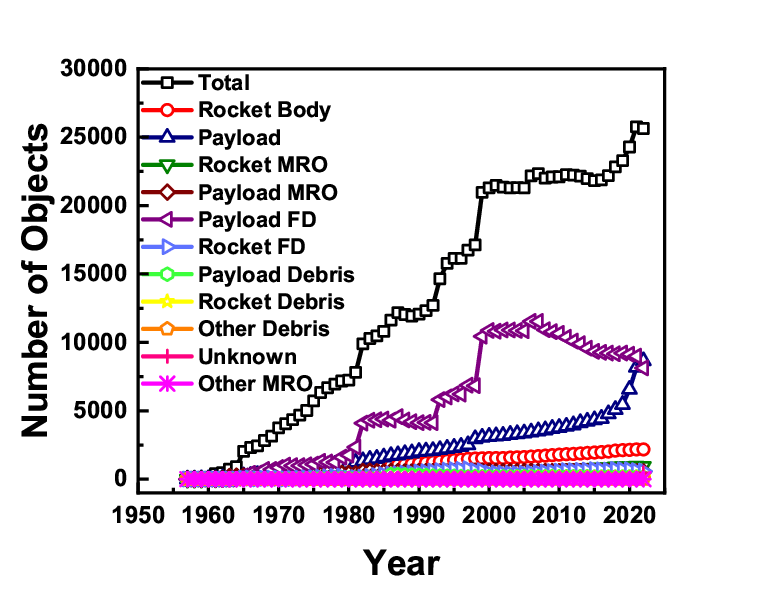}
    \caption{Monthly Number of Cataloged Objects in orbit. Here MRO and FD stand for "Mission related objects" and "Fragmentation Debris", respectively \cite{national}.}
    \label{fig:objects_on_orbit}
\end{figure}

\begin{figure}[h]
    \centering
    \includegraphics[width=0.9\linewidth]{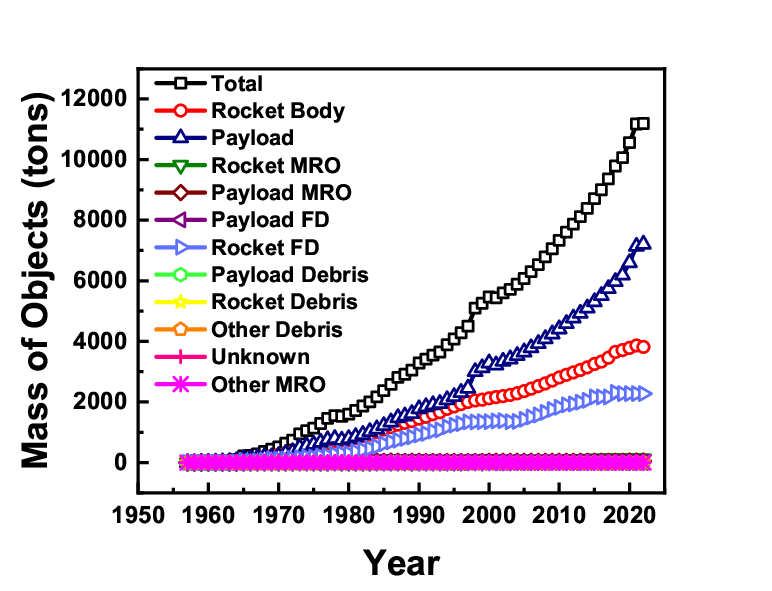}
    \caption{Monthly mass of objects in orbit based on object type until 2021 by NASA\cite{National_February}.}
    \label{fig:8}
\end{figure}

\begin{figure}[h]
    \centering
    \includegraphics[width=0.9\linewidth]{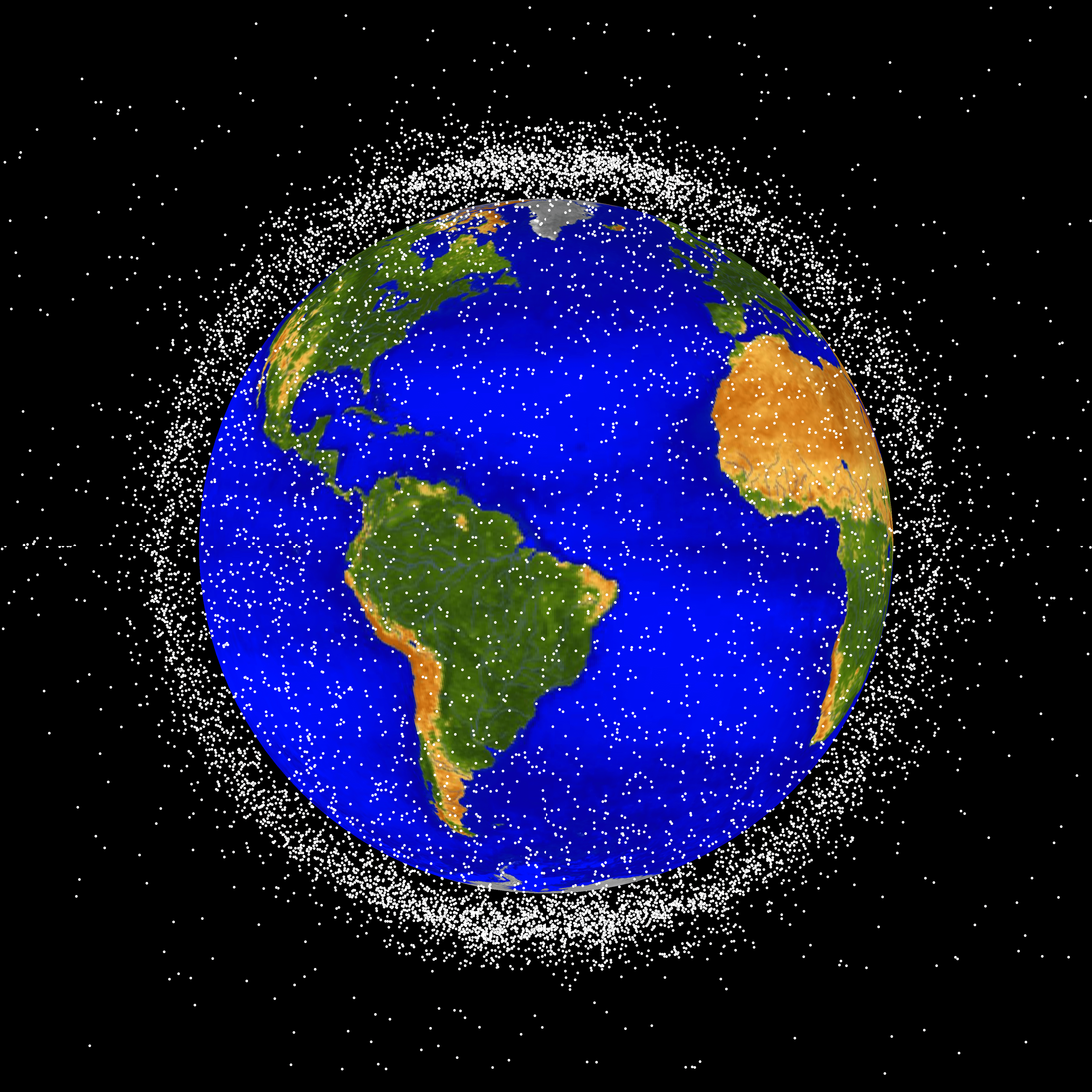}
    \caption{Illustration of objects present in the low Earth orbit (credit: NASA)\cite{aresNasa}.}
    \label{fig:earth_leo_debris}
\end{figure}

According to the recent NASA Orbital Debris Program Office (ODPO) report (March, 2022)\cite{nasaaa},\cite{national}, the monthly number
of cataloged objects ($>$ 10 cm) in LEO reaches around
25,500 in the year 2022 (see Fig.
\ref{fig:objects_on_orbit}). As shown in  Fig.
\ref{fig:objects_on_orbit}, payload and its fragmentation debris account
for the vast majority of the total number of cataloged objects. It is also seen from Fig.
\ref{fig:objects_on_orbit} that there has been a steep increase in payloads in recent years. 
\newacronym{esa}{ESA}{European Space Agency}
\newacronym{esoc}{ESOC}{European Space Operations Center}
NASA ODPO also reported the evolution of the mass of monthly cataloged objects from 1957 to 2021, presented in Fig. \ref{fig:8}. It has been seen from the figure that the total mass of monthly cataloged objects is around 11,000 tonnes, in which the contribution of payload, rocket bodies and its fragmentation debris are about 7,000, 4,000 and 2,000 tonnes, respectively\cite{National_February}.
\newacronym{nasa}{NASA}{National Aeronautics and Space Administration}
An illustration of the cataloged objects in low Earth orbit (see Fig. \ref{fig:earth_leo_debris}) is presented by NASA ODPO and they reported that about 95\% of the objects are orbital debris\cite{aresNasa}.

\newacronym{discos}{DISCOS}{Database and Information System Characterising Objects in Space}

\section{\large Design parameters for debris removal system}
\begin{figure}[h]
    \centering
    \includegraphics[width=0.9\linewidth]{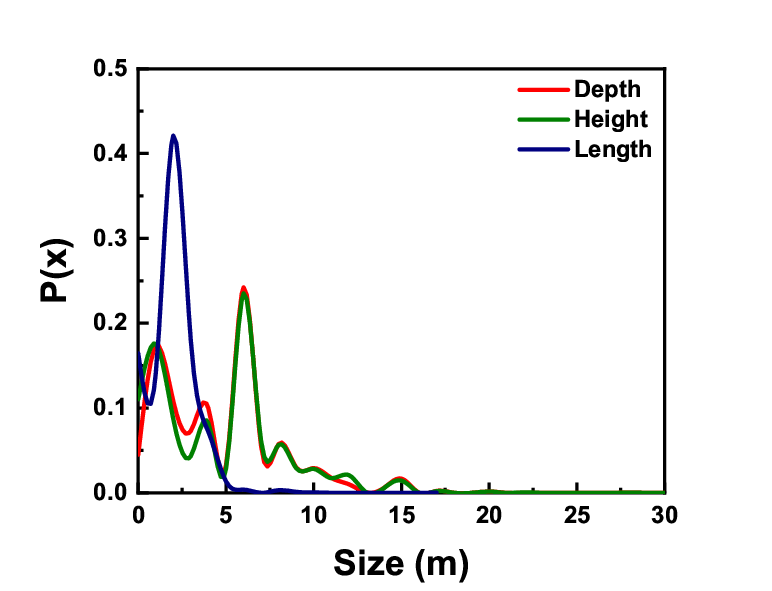}
    \caption{The probability distribution of size of debris in Earth's orbit.\cite{discosweb}}
    \label{fig:pdf_size}
\end{figure}

To design an efficient debris removal system, it is important to understand the physical properties (such as size, mass, and average area of cross-section) of the debris. The raw data related to physical properties can be found in the ESA's Database and Information System Characterising Objects in Space (\acrshort{discos}) \cite{discosweb}. The probability distribution of the size (depth, height, and diameter) of debris is present in Fig.\ref{fig:pdf_size}. Using the probability distribution function, we can analyze the distribution of debris size at different orientations. The PDF of height and depth shows a similar trend, with maximum probability at around 6 m and some smaller peaks around 1 m, 4 m, 9 m, and 15 m, whereas the probability distribution of length shows that the most probable length of debris is around 2 m. Thus, from the overall analysis of the PDFs of depth, height, and length, we can conclude that the most probable size of debris is around 2 m in length, 6 m in depth, and 6 m in height. The maximum probability of size with different orientations ranges between 0 and 7 m. It has also been observed (Fig. \ref{fig:pdf_size}) that the probability of large-sized debris (>20 m) is negligible. Thus, for the efficient design of debris removal systems, we need to focus on a system that can easily capture debris with a size of under 20 m. Extra attention is needed to design an efficient system that can capture smaller debris ranging from 0 to 7 m.

\begin{figure}[h]
    \centering
    \includegraphics[width=0.9\linewidth]{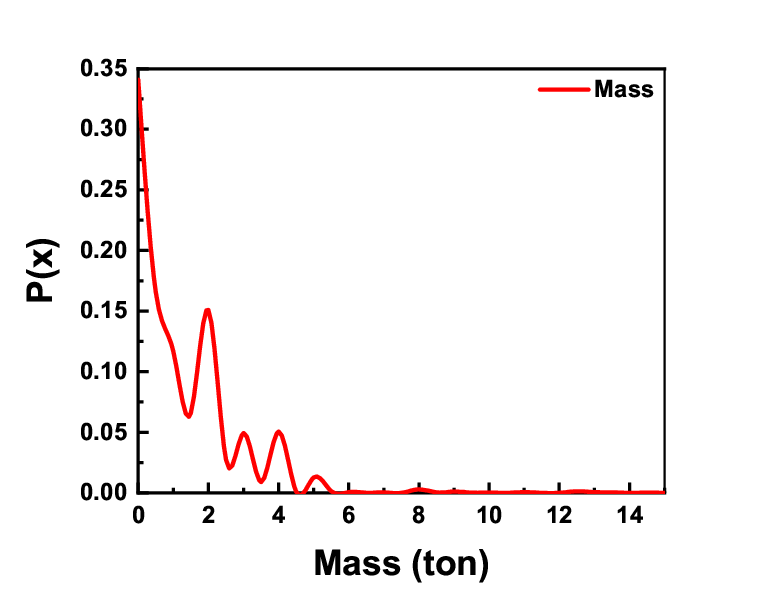}
    \caption{The probability distribution of mass of debris in Earth's orbit \cite{discosweb}.}
    \label{fig:PDF_mass}
\end{figure}

The distribution of mass of debris in the Earth's orbits is calculated using the database available on ESA's DISCOSweb and presented in Fig. \ref{fig:PDF_mass}. The probability distribution of mass shows that debris with a mass of less than 1 ton is more probable, with a decremented probability of 2, 4, 3, and 5 tons. The probability of a mass of debris greater than 6 tons is negligible. Thus, the debris capturing system must be designed, taking into consideration that debris with a mass of less than 1 ton is more probable, and it should be able to handle debris with a mass in the range of 0–6 tons.

\begin{figure}[h]
    \centering
    \includegraphics[width=0.9\linewidth]{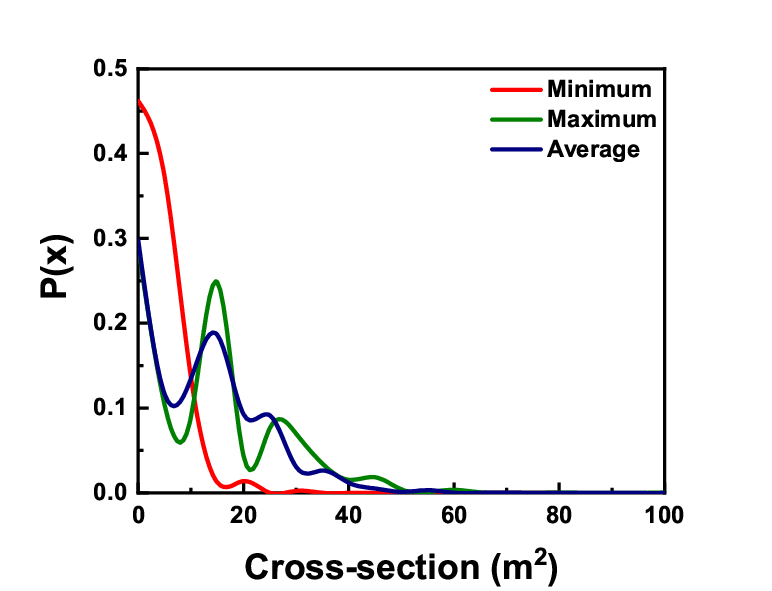}
    \caption{The probability distribution of area of cross-section of debris in Earth's orbit \cite{discosweb}}
    \label{fig:PDF_cross}
\end{figure}

\newacronym{pdf}{PDF}{Probability Distribution Function}
The cross-sectional area of the debris is also one of the important parameters for the debris capturing and removal system. Therefore, we analysed the probability distribution of the area cross-section of the debris in Earth's orbit and showed it in Fig. \ref{fig:PDF_cross}. Approximately 21,000 datasets are used to calculate the probability distribution function. Based on our results, the probability distribution of the cross-sectional area of debris suggests that debris with a cross-sectional area of less than 10 m$^2$ is most probable. However, we also found that the PDF analyzed peaks near 18 and 28 m$^2$. The probability is diminishing for large debris exceeding a cross-sectional area of 50 m$^2$. According to our analysis, the debris capturing system should be able to seize debris with a cross-sectional area of less than 50 m$^2$, with a particular emphasis on the debris with cross-sectional area of less than 10.0 m$^2$.

\begin{figure}[h]
    \centering
    \includegraphics[width=0.9\linewidth]{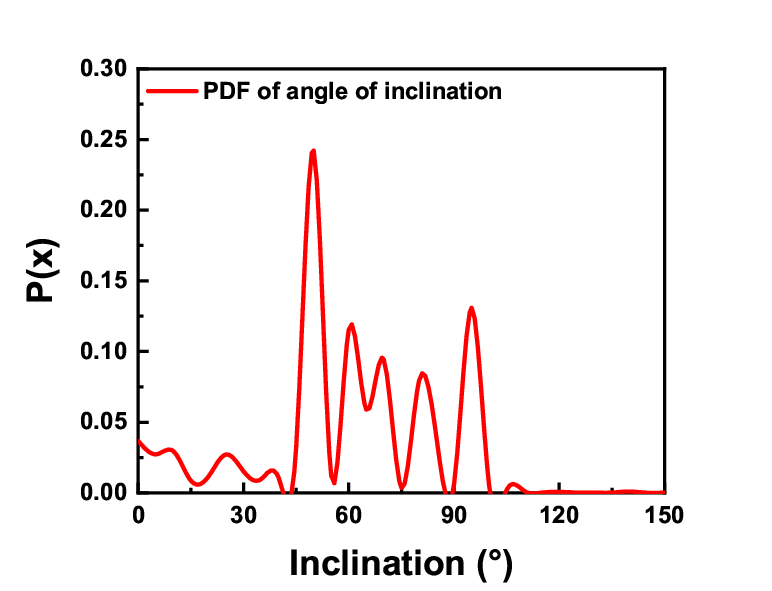}
    \caption{The probability distribution of the angle of inclination of objects in  Earth's orbit\cite{space-track}.}
    \label{fig:PDF_inclination}
\end{figure}

Figure \ref{fig:PDF_inclination} depicts the probability distribution of an object's angle of inclination in Earth's orbit. Based on the PDF of the inclination angle of the debris, it was found that the PDF represents most of the distribution of the inclination angle of the debris around 45 to 105 degrees, with a peak at about 50 degrees, demonstrating it as a most probable angle of inclination.                

\begin{figure}[h]
    \centering
    \includegraphics[width=0.9\linewidth]{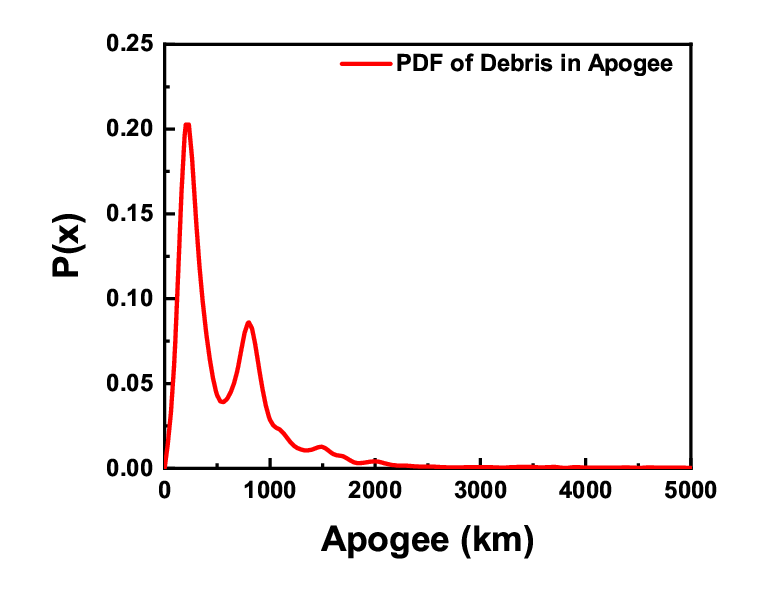}
    \caption{The probability distribution of debris position in apogee.}
    \label{fig:PDF_Apogee}
\end{figure}

\begin{figure}[h]
    \centering
    \includegraphics[width=0.9\linewidth]{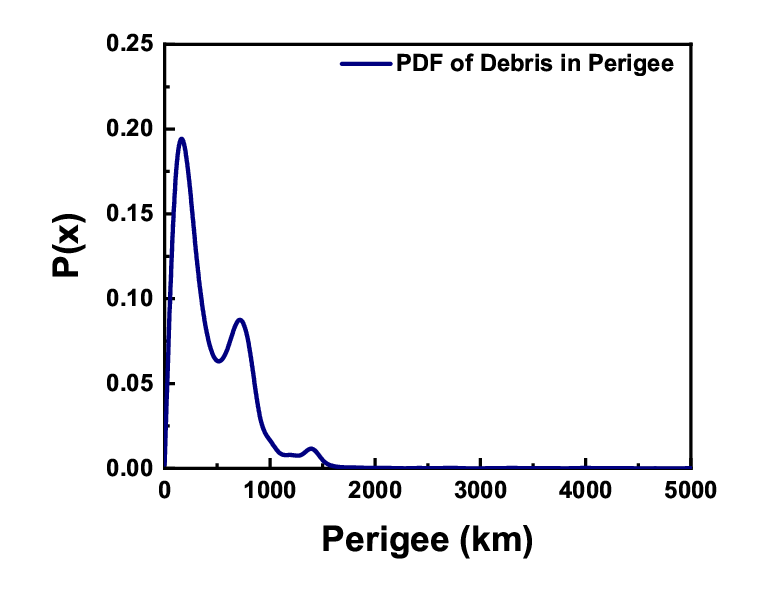}
    \caption{The probability distribution of debris position in perigee.}
    \label{fig:PDF_perigee}
\end{figure}

\begin{figure}[h]
    \centering
    \includegraphics[width=0.9\linewidth]{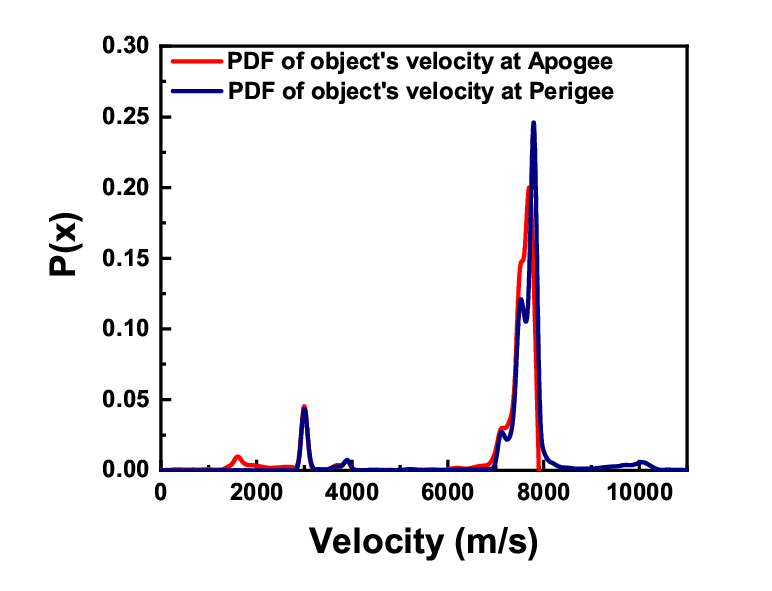}
    \caption{The probability of velocity of object in  Earth's orbit.}
    \label{fig:PDF_velocity}
\end{figure}

Figs. \ref{fig:PDF_Apogee}, \ref{fig:PDF_perigee} show PDF plots of the debris position at apogee and perigee in kilometers. According to our results, the probability distribution of the debris position ranges from 0 to 2000 km at apogee and from 0 to 1500 km at perigee. Most of the debris was found around 200 km from both the apogee and perigee, with the second most probable region for debris being around 800 km and 700 km from the apogee and perigee, respectively. For computing these probability distribution functions, around 32000 data sets were considered. Furthermore, the probability distribution of the velocity of the object at apogee and perigee is presented in Fig. \ref{fig:PDF_velocity}. The velocity distribution of objects at apogee and perigee shows that the most probable velocities of the debris at apogee and perigee are 7700 m/s and 7800 m/s, respectively. It has also been observed that a velocity of 3000 m/s is the second most probable velocity in both apogee and perigee. Overall, from the PDF plots, it has been seen that the probabilities of velocities at apogee and perigee are approximately the same.

\begin{figure}[h]
    \centering
    \includegraphics[width=0.9\linewidth]{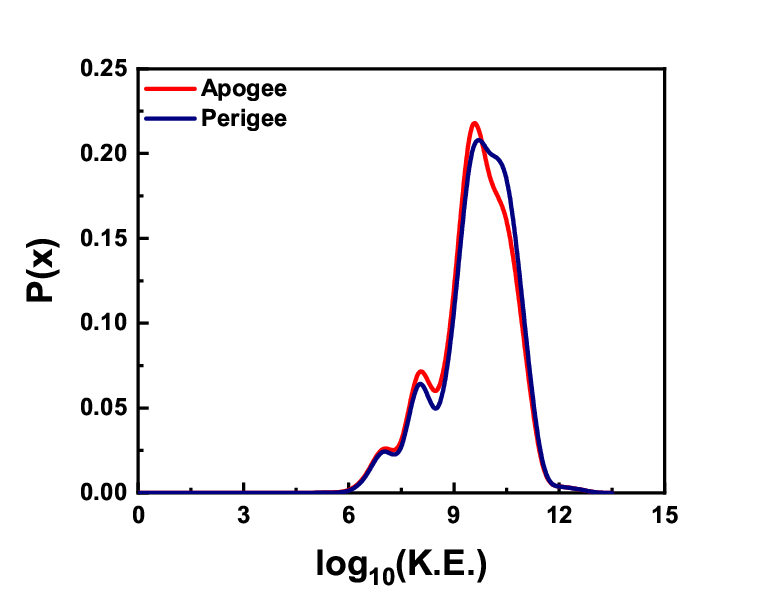}
    \caption{The probability distribution of kinetic energy (in Joules) of object at apogee and perigee.}
    \label{fig:PDF_kinetic_energy}
\end{figure}

In order to compute the kinetic energies (K.E.) of the objects at apogee and perigee, information related to the mass and altitude of the objects at apogee and perigee is needed. This information is present in two different databases, i.e., in ESA DISCOSweb and space-track. Therefore, the databases were first merged based on the cataloged ID. Then the kinetic energies (in joules) were computed by using velocities and masses obtained from the merged database. Further, the probability distribution function of kinetic energies at apogee and perigee are computed on a logarithmic scale ($log_10$). The probability distribution of the kinetic energy of objects at the apogee and perigee is presented in Fig .\ref{fig:PDF_kinetic_energy}.
The most probable kinetic energies at the apogee and perigee range from around $10^9$ to $10^{10}$ Joules.

\begin{figure}[h]
    \centering
    \includegraphics[width=0.9\linewidth]{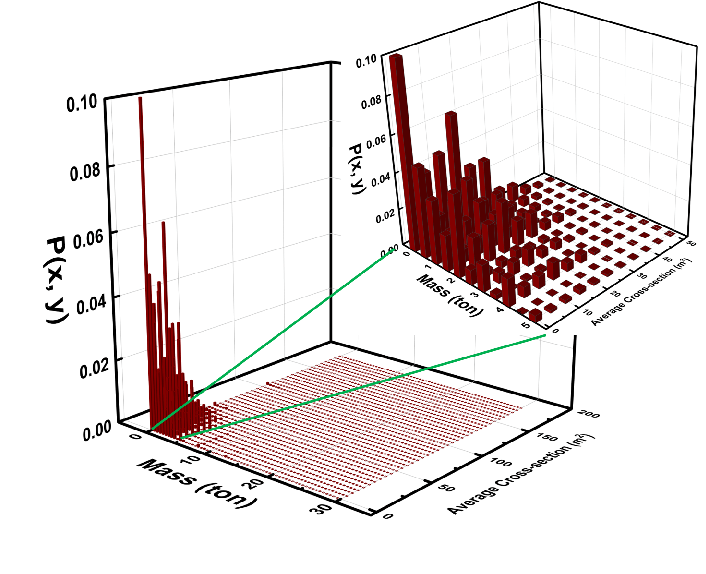}
    \caption{The joint probability distribution of mass and average area of cross-section of debris in Earth's orbit\cite{discosweb}.}
    \label{fig:PDF_joint_mass_cross}
\end{figure}

The joint probability of mass and average area of cross-section of debris is presented in Fig. \ref{fig:PDF_joint_mass_cross}. According to the plot, the most probable distributions of debris for the mass and average cross-section are around 0-5 tons and 0-40 m$^2$, respectively. 

The parameters discussed above, including size, mass, cross-sectional area, altitude, inclination, and kinetic energies of catalogued debris, will be useful for the design and development of an efficient debris removal system. 

\section{\large Mitigation Strategies}

Constraining the growth of space debris in order to preserve Earth's orbital space, particularly LEO, is an urgent challenge for space agencies \cite{nasa111,ellery2019tutorial}. The launch activities and operational practices have resulted in increased amount of debris in the Earth's orbits, creating a problem for other functional satellites and spacecraft. Measures such as de-orbiting non-functional satellites should be taken to reduce collision rates and constrain the amount of debris \cite{KLINKRAD20041251}. The development of analytical and numerical models for high-speed collision testing are a good way to reduce potential risk of damage and fragmentation of satellites. The inclusion of parameters such as velocity, mass, shape, size, and area of cross-section of debris in these models allows us to produce more accurate and useful information for testing damage mitigation techniques\cite{national1995orbital}. These numerical and analytical models would be helpful for the design and development of protective shields for satellites.

The methods to reduce the growth of space debris can be divided into two categories: the short-term risk reduction method and the long-term risk reduction method\cite{national1995orbital}. Short-term risk reduction methods involve collision avoidance maneuvers that change the path of a satellite, reducing the risk of collisions and the creation of further debris. However, it does not reduce the long-term risk of collisions. The long-term risk reduction methods include the de-orbiting of non-operational satellites by force entry into the Earth's atmosphere, the active removal of space debris, and the transfer of objects or debris into less populated orbits \cite{mcknigh1991determination,national1995orbital}.

\newacronym{das}{DAS}{Drag augmentation Systems}
Numerous collision risk reduction studies and space debris mitigation strategies have been explored, such as de-orbiting satellites from LEO. They are utilizing the drag augmentation deployable drag sail\cite{VISAGIE201565}, or electromagnetic tethering \cite{caldwell_2021}. Drag augmentation systems (\acrshort{das}) reduce the de-orbit period of a satellite by the drag sailing using stored energy for deployment. Thereby the likelihood of collisions is reduced\cite{SERFONTEIN2021278,serfontein2021drag}.
\newacronym{roger}{ROGER}{Robotic geostationary orbit restorer}

 Space agencies and companies are involved in addressing the capturing mechanism of debris\cite{li2021recent} and some of the concepts are: the flexible design of the "e.Deorbit" project of ESA \cite{telaar2017coupled}, the "Robotic geostationary orbit restorer" (\acrshort{roger}) \cite{bischof2003roger}, and the "Junk Hunter" program proposed by JAXA \cite{zinner2011junk}.

The net capturing method is one of the most promising methods for the captivity of debris because of its numerous advantages. For example, this method requires lower accuracy in the position and size of the debris. As nets are made up of elastic materials, they can be designed to cover a large surface areas, non uniform objects. Nets can be deployed towards a distant target using a firing mechanism that can capture debris in a cost-effective manner \cite{biesbroek2012deorbit}. Other advantages of net based capturing method include light weight, adaptation to the shape of the debris after capture, and the ability of the pursuer to maintain a safe distance from the uncontrolled target\cite{BARNES2020455}. In the year 2018, the in-orbit demonstration mission "RemoveDEBRIS" based on net capturing method successfully captured the cubesat showing its potential capacity\cite{AGLIETTI2020310,BARNES2020455}. Thus, the next sub-section focuses on the dynamics of the net capturing system due to various advantages.

\subsection{\large Modeling of Dynamics of the Net Capturing System}

\begin{figure}[h]
    \centering
    \includegraphics[width=0.9\linewidth]{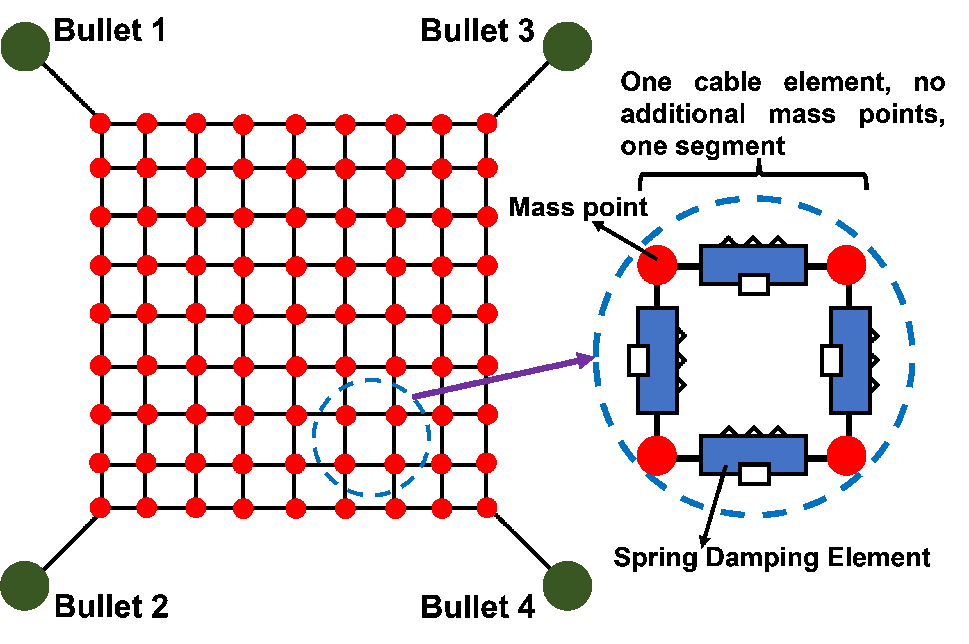}
    \caption{Mass-spring model. "Adapted from Acta Astronautica, Vol 132, Minghe Shan and Jian Guo and Eberhard Gill, Deployment dynamics of tethered-net for space debris removal, 293-302, Copyright (2017), with permission from Elsevier"
    \cite{SHAN2017293}.}
    \label{fig:Mass-spring model}
\end{figure}

There are several methods for modeling flexible nets, including the mass-spring model and  the absolute nodal coordinate formulation (ANCF) model\cite{SHAN20201083}.
In the mass-spring model, the net model is represented by discretizing cables into spring-mass-damper elements as shown in Figure \ref{fig:Mass-spring model}  \cite{SHAN20201083}. For the flexibility of the net, the spring element can be inserted between two pairs of mass points. The flexibility of the cable can be increased by adding several mass-spring segments between the two nodes. The addition of more mass points not only increases the freedom of movement of the cable but also the complexity of the simulation\cite{stadnyk2020validating}. The deployment of the net is performed by shooting four bullets (mass points) attached to the corners of the net\cite{SHAN20201083}.

\begin{figure}[h]
    \centering
    \includegraphics[width=0.9\linewidth]{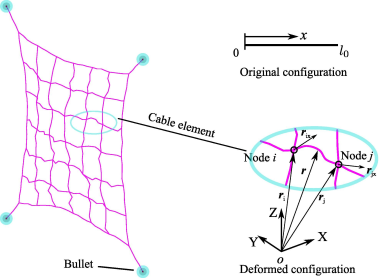}
    \caption{Net modeling based on the ANCF model. "Reprinted from Advances in Space Research, Vol 65, Issue 3, Minghe Shan and Jian Guo and Eberhard Gill, An analysis of the flexibility modeling of a net for space debris removal, 1083-1094, Copyright (2020), with permission from Elsevier"
    \cite{SHAN20201083}.}
    \label{fig:ANCF_Method}
\end{figure}

\newacronym{ancf}{ANCF}{Absolute nodal coordinate
formulation}
Multi-body problems with large displacements and deformations can be analyzed by the \acrshort{ancf} method (a nonlinear finite element formulation \cite{shabana2020dynamics}. The ANCF model can be used to  describe the flexibility of the net with fewer elements\cite{recuero2016chrono,huang2018review}. As shown in Fig \ref{fig:ANCF_Method}, there is one element (cable) between the two nodes\cite{SHAN20201083}.
The arbitrary position on the deformed cable element can be expressed  using the shape function $\textbf{\textit{S}}$ and the coordinates of the two nodes as \cite{recuero2016chrono,kim2018design}:

\begin{equation}
    \label{eq:1}
    \mathbf{r} = \textbf{\textit{Se}}= [S_1\textbf{\textit{I}}, S_2\textbf{\textit{I}}, S_3\textbf{\textit{I}}, S_4\textbf{\textit{I}}][\textbf{\textit{e}}_1, \textbf{\textit{e}}_2]^{\text{T}}
\end{equation}

where $\textbf{\textit{I}}$ is a $3 \times 3$ identity matrix and $\textbf{\textit{e}}_i$ is the absolute nodal coordinates\cite{recuero2016chrono,kim2018design}.

In terms of nodal coordinates (see Eq.\ref{eq:2}), $\textbf{\textit{r}}_j$ and $\textbf{\textit{r}}_{[j,x]}$ are the element's global displacements and slopes\cite{recuero2016chrono,kim2018design,SHAN20201083}. 

\begin{equation}\label{eq:2}
    \textbf{\textit{e}}_j= [\textbf{\textit{r}}_j, \textbf{\textit{r}}_{j,x}]
\end{equation}

The shape functions $S_i$ are defined as,

\begin{equation}\label{eq:3}
\begin{split}
    S_1 & = 1-3\xi^2 + 2\xi^3\\
    S_2 & = l_0(\xi - 2\xi^2 + \xi^3)\\
    S_3 & = 3\xi^2 - 2\xi^3\\
    S_4 & = l_0(-\xi^2 + \xi^3)
\end{split}
\end{equation}
 where $\xi = x/l_0$, and x is the coordinate of any point on the element\cite{recuero2016chrono,kim2018design,SHAN20201083}.

In order to derive the system equations of motion for the tethered net, one must use the principle of virtual power and introduce Lagrange multipliers \cite{SHAN20201083}.

\begin{equation}\label{eq:4}
\begin{bmatrix}
  \textbf{\textit{M}}_{b} & \textbf{\textit{0}} & \boldsymbol{\Phi}_{b}^\text{T} \\
  \textbf{\textit{0}}   & \textbf{\textit{M}}_{e} & \boldsymbol{\Phi}_{e}^\text{T}\\
  \boldsymbol{\Phi}_{b} & \boldsymbol{\Phi}_{e} & \textbf{\textit{0}}
\end{bmatrix}
\begin{bmatrix}
\boldsymbol{\ddot{q}}_{b}\\
\boldsymbol{\ddot{e}} \\
\lambda
\end{bmatrix}
=
\begin{bmatrix}
\textbf{\textit{Q}}_b\\
\textbf{\textit{Q}}_e\\
\textbf{\textit{Q}}
\end{bmatrix}
\end{equation}

where the mass matrix of bullets and the external forces on the bullets are given by $\textbf{\textit{M}}_{b}$ and $\textbf{\textit{Q}}_b$, respectively. The term $\textbf{\textit{M}}_{e}$ corresponds to a constant ANCF mass matrix, and $\textbf{\textit{Q}}_e$ is the generalized force associated with the absolute nodal coordinates $\boldsymbol{e}$. Here $\textbf{\textit{Q}}_e$ represents the external forces plus the elastic forces. The Lagrange multipliers are defined by $\boldsymbol{\lambda}$; the Jacobi matrix of the constraint associated with the absolute nodal coordinates is $\boldsymbol{\Phi_{e}}$; $ \boldsymbol{\Phi}_{b}$ represents the constraints coupled by the bullet masses and the cable elements; $\boldsymbol{Q}$ is a quadratic velocity vector\cite{SHAN20201083,shan2017deployment}.

The Kelvin-Voigt model \cite{eringen1980mechanics} is an efficient model for computing the dynamics of flexible nets, and is sufficiently accurate for simulating large-scale flexible nets \cite{ru2022capture}. Because of the inherent characteristics of cable material, cable elements are unable to sustain the compression and tension forces are generated, only when the cables are elongated. According to the Kelvin-Voigt method, the tension force $\textbf{\textit{f}}_{ij}$ between nodes \textit{i} and \textit{j} in a flexible net can be described as follows\cite{SHAN2019198},

\begin{equation}
\textbf{\textit{f}}_{ij}=
\\
\begin{cases}
\textit{f}_{ij} \widehat\textbf{\textit{r}}_{ij} \qquad \textit{r}_{ij} >\textit{l}_{0}\qquad and\qquad \textit{f}_{ij} <0,

\\ 
0 \qquad\qquad \textit{r}_{ij} \leq \textit{l}_{0}\qquad or\qquad \textit{f}_{ij} > 0,
\end{cases}   
\end{equation}

$\widehat\textbf{\textit{r}}_{ij}$ is the unit direction vector of the two adjacent nodes of the segment\cite{SHAN2019198,ru2022capture}.
The $\textit{f}_{ij}$ is given by: 
$$\textit{f}_{ij} = \textit{-k}(\textit{r}_{ij} - \textit{l}_{0})-\textit{c}\dot {\textit{r}_{ij}}$$
where the relative displacement and relative velocity between the $\textit{i}^{\text{th}}$ and $\textit{j}^{\text{th}}$ nodes are $\textit{r}_{ij}$ and $\dot{\textit{r}_{ij}}$, respectively. The stiffness constant of a net cable is defined by  $\textit{k}= \textit{EA} /\textit{l}_0$, where ($\textit{E}$) is  the modulus of elasticity, which can vary from material to material, $\textit{A}$ is the cross-sectional area of the cable, and $\textit{l}_0$ the initial length \cite{SHAN2019198,ru2022capture}. The damping coefficient of the cable material $c$ depends on the stiffness $k$ as 
\begin{equation} \label{eq:6}
\textit{c}_i = 2\xi \sqrt{\textit{m}_i \textit{k}}
\end{equation}
 where $\xi$ is the damping ratio, and $\textit{m}_i$  is the $\textit{i}^{\text{th}}$ lumped mass.

\begin{figure}[h]
    \centering
    \includegraphics[width=0.9\linewidth]{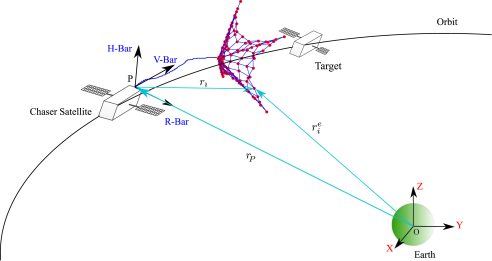}
    \caption{The chaser satellite is approaching the target in a track-along direction (V-Bar). "Reprinted from Acta Astronautica, Vol 158, Minghe Shan and Jian Guo and Eberhard Gill, Contact dynamic models of space debris capturing using a net, 198-205, Copyright (2019), with permission from Elsevier" \cite{{SHAN2019198}}.}
    \label{fig:chaser satellite}
\end{figure}

Before launching the net, the chaser satellite is supposed to move toward the target in an along-track direction \cite{shan2018contact,SHAN2019198}, as depicted in Fig. \ref{fig:chaser satellite}. 
The inertial reference frame for the net capture dynamics is centered on the Earth, and the local reference frame is centered on the center of mass of the net launch system \cite{shan2018contact,SHAN2019198}. The absolute position of mass points in the inertial reference frame can be expressed as\cite{SHAN2019198},
 
\begin{equation}\label{eq:7}
\textbf{\textit{r}}_{i}^{e} = \textbf{\textit{r}}_{P} + \textbf{\textit{R}}^{o} \textbf{\textit{r}}_{i}
\end{equation}

where in the inertial frame, $\textbf{\textit{R}}^{o}$ is the rotation matrix which transform the position vector $\textbf{\textit{r}}_{i}$ from the local system into the inertial system and $\textbf{\textit{r}}_{i}^{e}$ represents the absolute position of the $\textit{i}^{\text{th}}$ node. Finally, the discretized equation of motion is expressed as follows\cite{shan2018contact,SHAN2019198},

\begin{equation}\label{eq:8}
 \textit{m}_{i}\frac{\textit{d}\dot{\textbf{\textit{r}}_{i}^{e}}}{\textit{dt}}=\sum_{j=1}^{N_i}\textbf{\textit{R}}^{o}\textbf{\textit{f}}_{ij} + \sum_{s=1}^{M_i} \textbf{\textit{f}}_{is}^{e} + \textbf{\textit{g}}_{i},
\end{equation}

In Eq. \ref{eq:8}, $N_i$ is the number of neighboring cables linked to the $i^{\text{th}}$ node, $\textbf{\textit{f}}_{ij}$ is the force acting on the $i^{\text{th}}$ 
node by the $j^{\text{th}}$ connected cable;  $M_i$ represents the number of external forces on the $i^\text{th}$ node. The sum of external forces, such as aerodynamic drag or solar radiation pressure, is represented by $\textbf{\textit{f}}_{is}^{e}$. The micro-gravitational force $\textbf{\textit{g}}_{i}$ on the $i^\text{th}$ node is given by\cite{shan2018contact,SHAN2019198},

\begin{equation}\label{eq:9}
  \textbf{\textit{g}}_{i}=-\frac{\textit{GMm}_i \textbf{\textit{r}}_{i}^{e}}{\lvert \textbf{\textit{r}}_{i}^{e} \rvert^3}
  ,
\end{equation}

where, $\textit{GM}$ is the gravitational coefficient of the Earth \cite{shan2018contact,SHAN2019198}.

\subsection{\large Contact Dynamics of the Capture System}

\begin{figure}[h]
    \centering
    \includegraphics[width=0.9\linewidth]{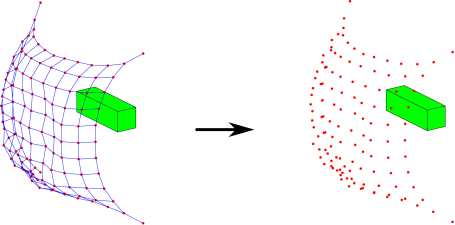}
    \caption{Net contact with an object on the left side and a simplified contact scenario on the right side. "Reprinted from Acta Astronautica, Vol 158, Minghe Shan and Jian Guo and Eberhard Gill, Contact dynamic models of space debris capturing using a net, 198-205, Copyright (2019), with permission from Elsevier"\cite{SHAN2019198}.}
    \label{fig:Net contacting}
\end{figure}

The dynamics of the capture system consist of the net capture dynamics and the debris dynamics. The net capture dynamics include the constitutive dynamics of the flexible net and its nonlinear contact dynamics with the debris. The debris dynamics include the translational and rotational dynamics of the debris\cite{ru2022capture}. As shown in Fig. \ref{fig:Net contacting}, when the net contacts a target, multiple mass points are in contact with the target surface because multiple mass points are connected to spring-damper elements in a specific pattern\cite{ru2022capture}. The number and position of the contacting points determine the response to a contact \cite{shan2018contact}.

Depending on the contact modeling method, the contact response is either penalty-based or impulse-based. In the penalty-based method, the contact force acting on the contact point is calculated as soon as contact is detected. In the impulse-based method, the contact impulse and velocity change are calculated. Then the modified velocities are subsequently incorporated into the dynamic equations of the net. However, contact detection is the same for both models \cite{shan2018contact,SHAN2019198}.

\subsubsection{\large Penalty-Based Contact Method}
In net-contact dynamics, a net is modeled as multiple mass points linked with spring-damper elements in a specific pattern. Therefore, net contact with a target can be thought of as multiple mass points contacting the target's surfaces. The mass points are considered to move outside the space of the target and are forbidden to penetrate it. Thus, according to this constraint, net capturing can be assumed to be an inequality constraint problem. The penalty-based method, which is basically used to solve optimization as well as inequality problems\cite{ru2022capture}. In the optimization problem, an inequality constraint can be expressed as \cite{SHAN20201083},

\begin{equation}
   \text{Minimize} \quad f(x), \quad \text{subject to  }  c(x) \geq 0
\end{equation}

where $f(x)$ and $c(x)$ are the objective function and constraint equation, respectively. The inequality constraint problem can be transformed into an unconstrained problem as \cite{SHAN20201083},

\begin{equation}
    \text{Minimize} \quad f(x) +\mu \: \text{min}\{0, c(x)\}, \quad \text{subject to  }x \in R^n
\end{equation}

where  $\mu \: \text{min}\{0, c(x)\}$ is defined as the penalty function.

If $c(x) \geq 0$, there is no penalty because  $\mu \: \text{min} \{0,c(x)\} = 0$ according to penalty function. In contrast, if $c(x) < 0$ , then $\mu \: \text{min} \{0,c(x)\} = c(x)$. The penalty is included in the objective function to prevent the penetration of a point in the target surface\cite{SHAN20201083}. When the point is not in contact with the target surface,  $c(x) \ge 0$, the constraint equation is  $\text{min} \{0,c(x)\}= 0$. However, when the point penetrates the target surface, $c(x)< 0$. The penalty force, here the contact force, $\mu c(x)$ is added to the objective function $f(x)$ to push the point away from the surface and prevent further penetration \cite{SHAN20201083}.
This definition considers the penalty function $\mu \: \text{min} \{0,c(x)\}$  to be a function of penetration depth. When two elastic bodies come into contact, the colliding bodies deform, and the contact force is expressed as a function of deformation. In simulations of a contact scenario, the deformation of the contact can be parameterized according to the depth of penetration between the colliding bodies. In the penalty-based method, the reaction force is calculated based on the depth of penetration: the deeper the penetration, the higher the penalty \cite{SHAN20201083}.

\subsubsection{\large Impulse-Based Contact Method}

In the impulse-based method, the impulse caused by the contact is calculated instead of the contact force. Thus, in this method, the change in velocity is computed after the contact \cite{shan2018contact}. Virtual reality and gaming environments typically use impulse-based methods for the contact between two separate bodies\cite{shan2018}. The basic idea of impulse-based simulation is that any contact between bodies is modeled by collisions at the contact points; non-penetration constraints do not exist; non-penetration constraints between different bodies are enforced by collisions \cite{mirtich1996impulse}. This method applies to contact dynamics in the removal of tumbling space debris using a net. The net is discretized into mass points that are connected in a specific pattern. Consequently, a net encountering a target can be considered a single mass point or multiple mass points of contact with the target object \cite{shan2018}.

\subsection{\large Results of Simulations and Experiments of the Space Debris Capture System}

\begin{figure}[h]
    \centering
    \includegraphics[width=0.9\linewidth]{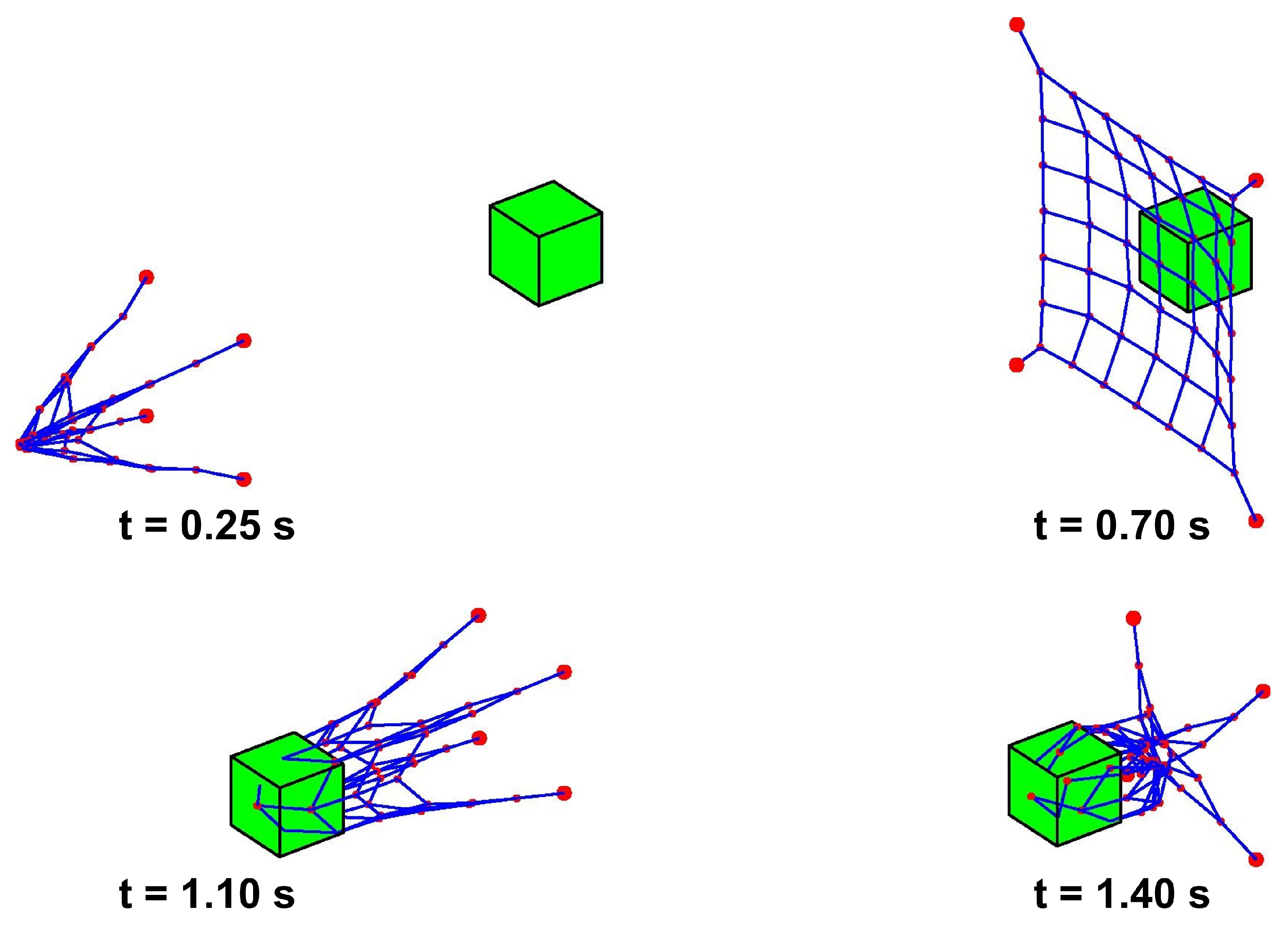}
    \caption{Capture analysis based on the mass-spring model. "Reprinted from Advances in Space Research, Vol 65, Issue 3, Minghe Shan and Jian Guo and Eberhard Gill, An analysis of the flexibility modeling of a net for space debris removal, 1083-1094, Copyright (2020), with permission from Elsevier"
    \cite{SHAN20201083}.}
    \label{fig:mass-spring}
\end{figure}

\begin{figure}[h]
    \centering
    \includegraphics[width=0.9\linewidth]{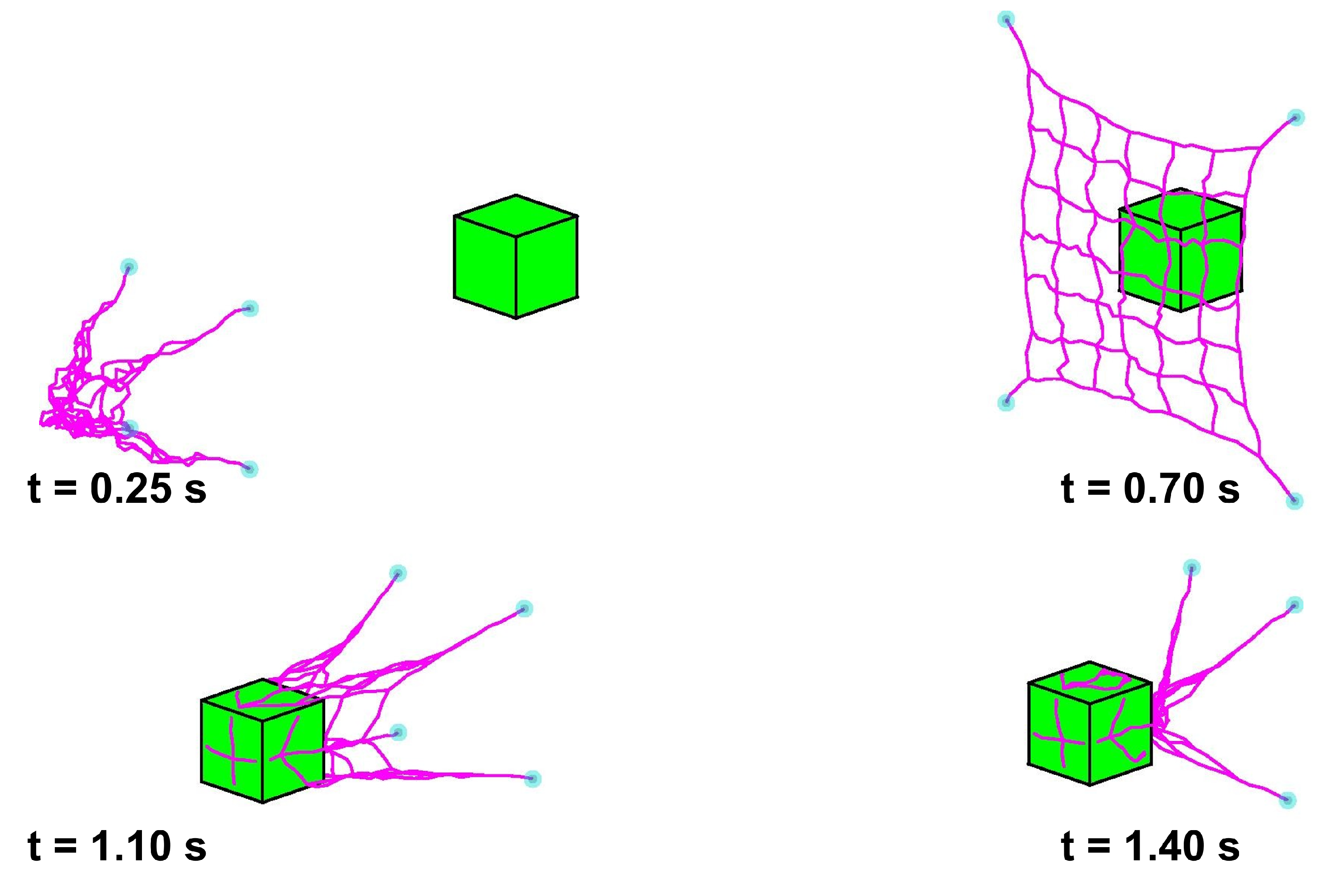}
    \caption{Capture analysis based on the ANCF model. "Reprinted from Advances in Space Research, Vol 65, Issue 3, Minghe Shan and Jian Guo and Eberhard Gill, An analysis of the flexibility modeling of a net for space debris removal, 1083-1094, Copyright (2020), with permission from Elsevier"
    \cite{SHAN20201083}.}
    \label{fig: ANCF model}
\end{figure}

Numerous numerical simulations were performed to study the dynamics of contact and capture of an object based on the mass spring (see Fig. \ref{fig:mass-spring}) and the ANCF (see Fig. \ref{fig: ANCF model}) models\cite{SHAN2019198,SHAN20201083,si2021dynamics}. To analyze and compare the effectiveness of the contact dynamic models, the penalty-based method and the impulse-based method were studied. The weaknesses and strengths of these contact dynamic models were determined\cite{SHAN2019198}. Based on the simulation results, both methods are found to be effective in modeling net contact dynamics\cite{huang2018review}.  

\begin{figure}[h]
    \centering
    \includegraphics[width=0.9\linewidth]{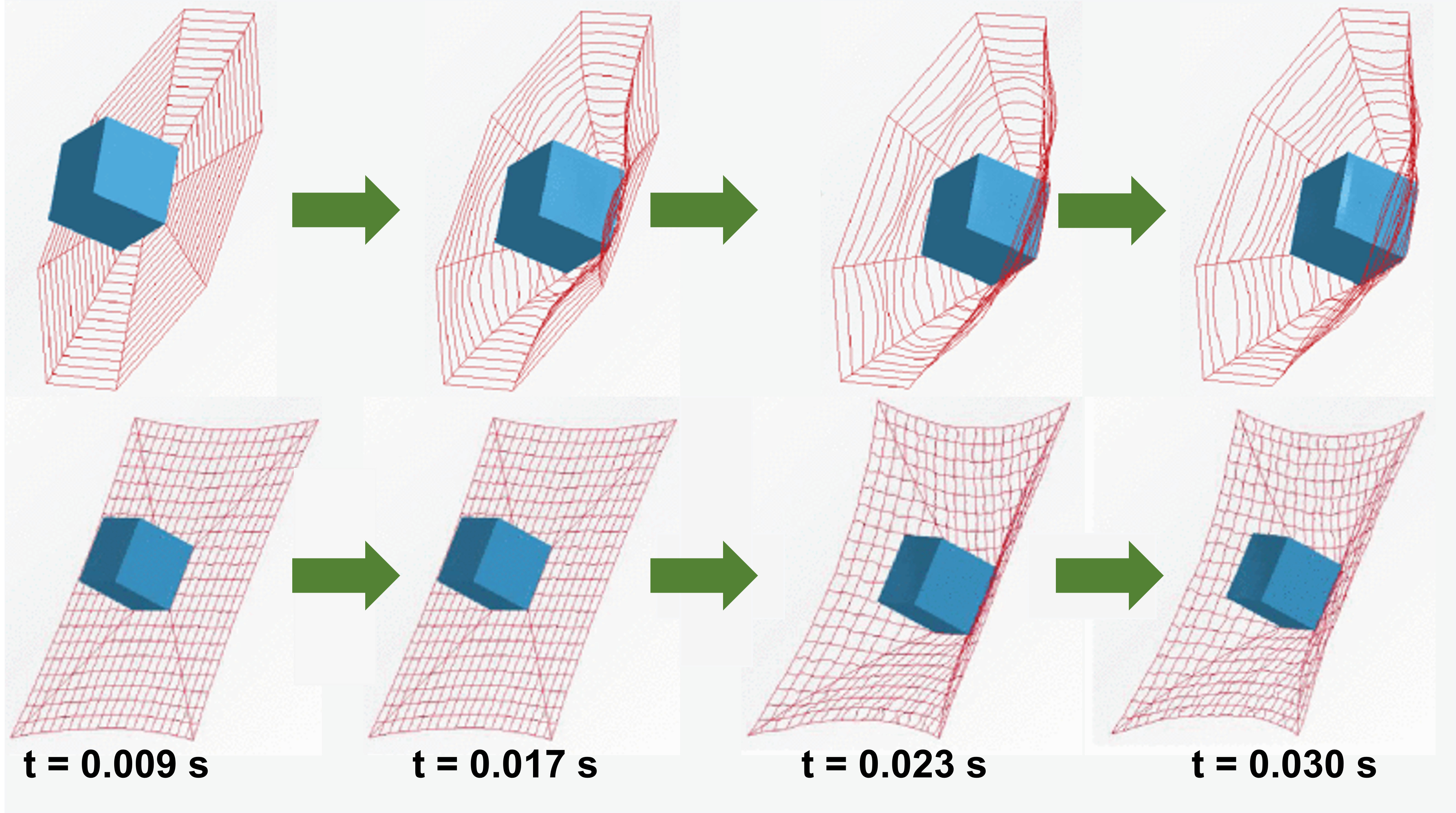}
    \caption{The deformation of two web configurations. ”Reprinted from in IEEE Access, Vol 8, Xu, Boting and Yang, Yueneng and Zhang, Bin and Yan, Ye and Yi, Zhiyong, Bionic Design and Experimental Study for the Space Flexible Webs Capture System, 45411-45420, Copyright (2020), Under Creative Commons License” \cite{xu2020bionic}.}
    \label{fig:web}
\end{figure}

In some studies, the structure of the debris-capturing system is inspired by nature and modeled as a the spider's web. The finite element method and experimental methods are used for collision analysis and designing a space debris-capturing system \cite{trivailo2016dynamics,xu2020bionic}. In a finite element simulation study, the collision of objects with the stationary web is simulated. Several cases were simulated and the impact of objects leading to failure or rupture of the web is reported \cite{trivailo2016dynamics}. The study also reported that spiral webs have better strength compared to radial webs.  Based on the dynamics of the simulations, it has been shown that a rotating web, unlike a non-rotating web, cannot completely capture objects. The wrapping of the objects depends strongly on their shape, their number, the relative speed of the web, the rotational angular velocity of the spin-stabilized web, and the direction of its spin \cite{trivailo2016dynamics}. In another study, the flexible web model has been compared with traditional quadrilateral webs. It has been observed that the bionic octagonal flexible webs have higher dissipation performance, high tolerance ability, and significant mechanical properties compared with the quadrilateral webs \cite{xu2020bionic}. The deformation of two web configurations is shown in Fig. \ref{fig:web} \cite{xu2020bionic}. 

\begin{figure}[h]
    \centering
    \includegraphics[width=0.9\linewidth]{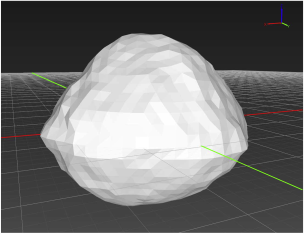}
    \caption{Asteroid Bennu 3D model with collision geometry. "Reprinted from Acta Astronautica, Vol 155, Eleonora M. Botta and Inna Sharf and Arun K. Misra, Simulation of tether-nets for capture of space debris and small asteroids, 448-461, Copyright (2019), with permission from Elsevier" 
    \cite{BOTTA2019448}.}
    \label{fig:Bennu}
\end{figure}

Botta et al. \cite{BOTTA2019448} presented a simulation study to gain insight into the dynamics of space debris capture. The major components of a simulator system includes the chaser, capture net, and closure mechanism.  In the simulator, a multi-body dynamics simulation platform  Vortex Dynamics was used to perform contact dynamics. Several models such as a standard lumped-parameter model, an augmented lumped parameter model, and a cable-based model of net are implemented in the  simulator, 
\begin{figure}[h]
    \centering
    \includegraphics[width=0.9\linewidth]{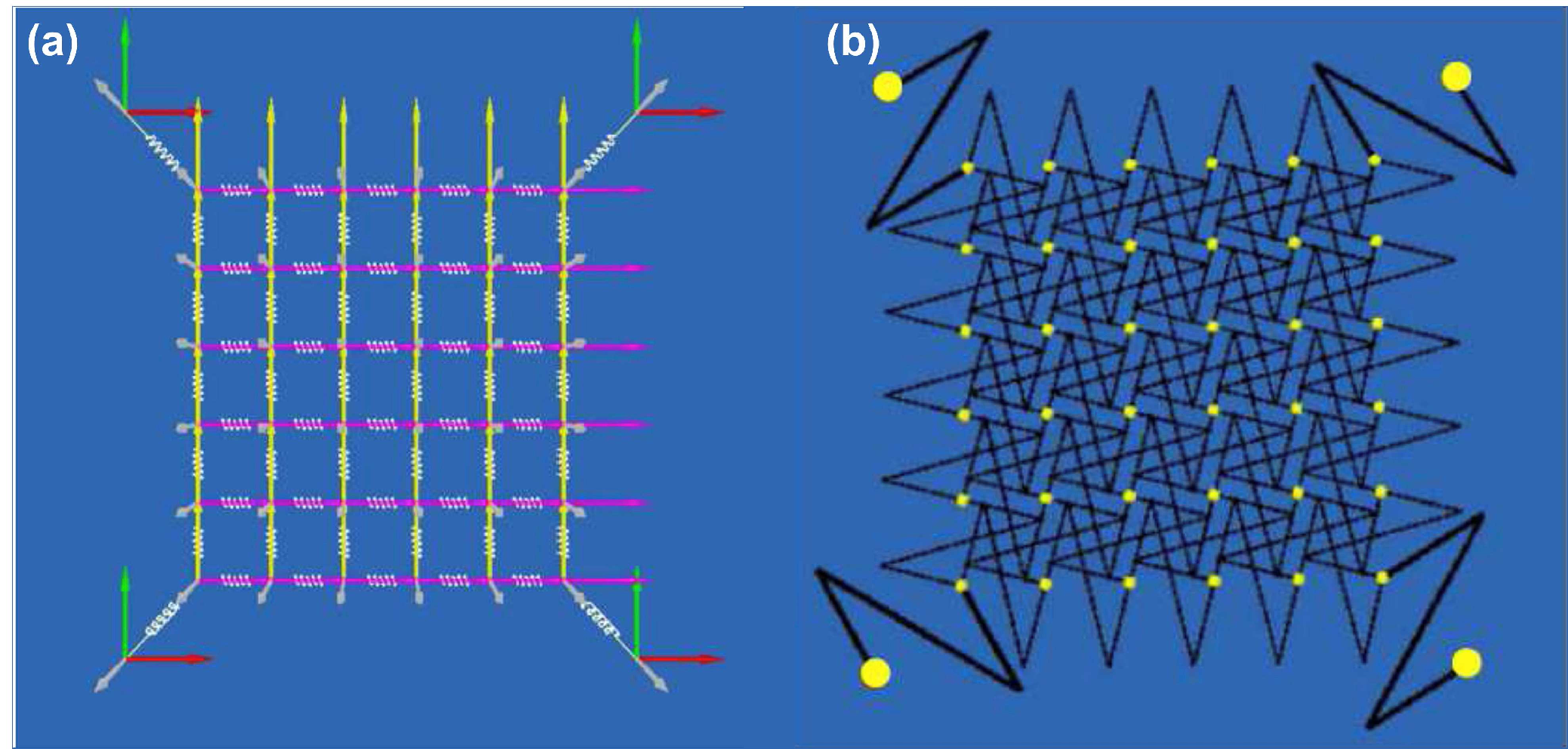}
    \caption{(a) The standard lumped-parameter, and (b) cable-based of the net models. "Reprinted from Acta Astronautica, Vol 155, Eleonora M. Botta and Inna Sharf and Arun K. Misra, Simulation of tether-nets for capture of space debris and small asteroids, 448-461, Copyright (2019), with permission from Elsevier" \cite{BOTTA2019448}.}
    \label{fig:lumped and cable modeles}
\end{figure}
The model with standard lumped parameters is shown in Fig. \ref{fig:lumped and cable modeles}(a). The mass properties of the net are represented by small spherical rigid bodies at the knots of the net, while the axial stiffness and damping properties are represented by mass-less springs and dampers between the small rigid bodies(nodes). In the lumped-parameter model, additional linear and torsional springs and dampers were added to describe the bending stiffness of the cables of the net. The cable-based model, which is a complex but more accurate model than the lumped-parameter model, is depicted in Fig. \ref{fig:lumped and cable modeles}(b). According to this model, the mass is dispersed through the net's threads, which are modeled as thin rigid bodies joined together at prismatic joints\cite{BOTTA2019448}. In this study, a library of realistic debris was created for Zenit-2 rocket stage 2, Envisat, Apollo spacecraft, and a small asteroid like Bennu. The asteroid Beenu is modeled as a free-floating, tumbling rigid body. The scaled down model of asteroid Bennu is shown in Fig.\ref{fig:Bennu}. Based on the scaled 3D model of Bennu, a convex mesh collision geometry was added \cite{BOTTA2019448}. 

\begin{figure}[h]
    \centering
    \includegraphics[width=0.9\linewidth]{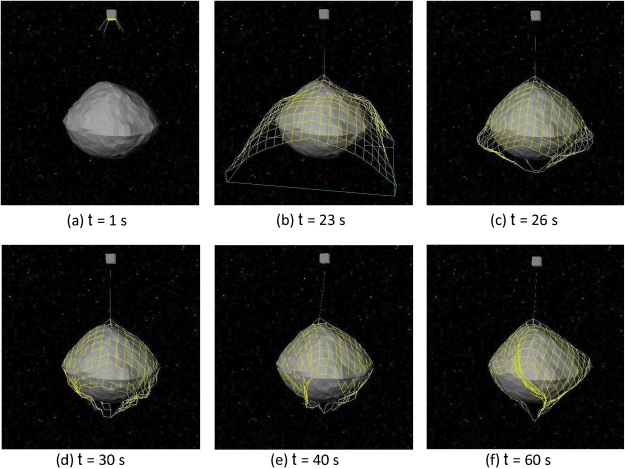}
    \caption{ Simulation of net capture with a standard closing mechanism: (a) deployment of the net, (b) activation of the closing mechanism, (c) achievement of closure, (d) capture of the asteroid, and (e)-(f) maintenance of capture. "Reprinted from Acta Astronautica, Vol 155, Eleonora M. Botta and Inna Sharf and Arun K. Misra, Simulation of tether-nets for capture of space debris and small asteroids, 448-461, Copyright (2019), with permission from Elsevier" \cite{BOTTA2019448}.}
    \label{fig:capture of Bennu}
\end{figure}

The snapshots of the capture sequence of asteroid Bennu are shown in Fig. \ref{fig:capture of Bennu}. At $t=23$ s (Fig.\ref{fig:capture of Bennu}(b)), the net has begun to capture, and enclose the asteroid (Fig.\ref{fig:capture of Bennu}(c)), whereupon the locking mechanism begins. Locking is almost achieved at $t=26$ s and is confirmed at $t=30$ s ((Figs.\ref{fig:capture of Bennu}(d)). As the simulation continues (Figs.\ref{fig:capture of Bennu}(e)-(f)), the 3D model of asteroid was observed trapped even after 60 seconds \cite{BOTTA2019448}.

\begin{figure}[h]
    \centering
    \includegraphics[width=0.9\linewidth]{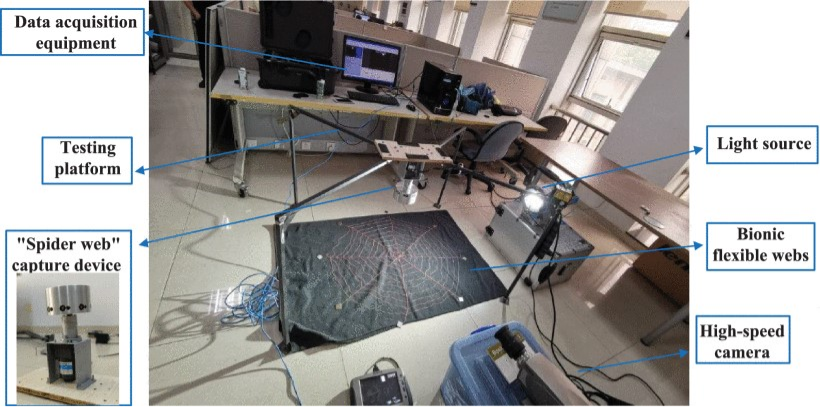}
    \caption{The ground test system of “cobweb-like” flexible webs capture device. ”Reprinted from in IEEE Access, Vol 8, Xu, Boting and Yang, Yueneng and Zhang, Bin and Yan, Ye and Yi, Zhiyong, Bionic Design and Experimental Study for the Space Flexible Webs Capture System, 45411-45420, Copyright (2020), Under Creative Commons License”\cite{xu2020bionic}.}
    \label{fig:spider_web}
\end{figure}

Experimental studies were also conducted to understand the debris capture mechanism using flexible nets. Fig.\ref{fig:spider_web} shows the ground verification experiment setup for the debris capture device. This device consists of flexible net with magnetic masses connected at corners, a central rigid body, a central shaft, transmission elements, a variable-speed motor and other components \cite{xu2020bionic}. The net is made up of polyethylene fibers in woven in octagonal shape. The polyethylene fiber segment has a diameter of 0.3 mm, and the diagonal diameter of the webs is 1.0 m. Magnetic iron is used to make the mass blocks that are attached to the corners of the net, and each magnetic mass weighs five grams\cite{xu2020bionic}.
\begin{figure}[h]
    \centering
    \includegraphics[width=0.9\linewidth]{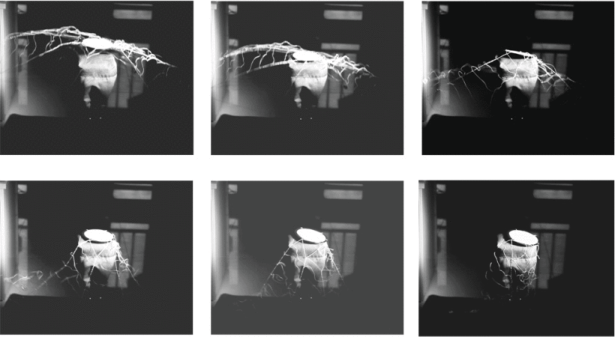}
    \caption{Collision of flexible webs with the target body. ”Reprinted from in IEEE Access, Vol 8, Xu, Boting and Yang, Yueneng and Zhang, Bin and Yan, Ye and Yi, Zhiyong, Bionic Design and Experimental Study for the Space Flexible Webs Capture System, 45411-45420, Copyright (2020), Under Creative Commons License” \cite{xu2020bionic}.}
    \label{fig:internal structure}
\end{figure}
The collision process between the flexible orbits and the target is presented in Fig.\ref{fig:internal structure}. The results show that the flexible web collided with the target body and wrapped it. Due to the presence of gravity, the shell nosing condition is slightly different on Earth, and ground testing also reported the weaving defects of the flexible web. Overall, the authors concluded that the bionic-designed (cobweb-like) flexible capture device can be an effective debris-capture system  \cite{xu2020bionic}.

\begin{figure}[h]
    \centering
    \includegraphics[width=0.9\linewidth]{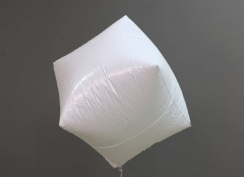}
    \caption{A cubic blimp is used as a debris for testing. "Reprinted from Acta Astronautica, Vol 139, Inna Sharf and Benjamin Thomsen and Eleonora M. Botta and Arun K. Misra, Experiments and simulation of a net closing mechanism for tether-net capture of space debris, 332-343, Copyright (2017), with permission from Elsevier" \cite{sharf2017experiments}.}
    \label{fig:helium blimp}
\end{figure}

\begin{figure}[h]
    \centering
    \includegraphics[width=0.9\linewidth]{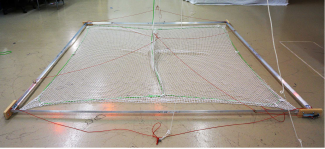}
    \caption{Net capture system attached to the frame. "Reprinted from Acta Astronautica, Vol 139, Inna Sharf and Benjamin Thomsen and Eleonora M. Botta and Arun K. Misra, Experiments and simulation of a net closing mechanism for tether-net capture of space debris, 332-343, Copyright (2017), with permission from Elsevier"  \cite{sharf2017experiments}.}
    \label{fig:Net-test}
\end{figure}

In another experiment for testing a debris-capturing system, a test bed is established in a laboratory with a height of 4 m from the floor to the ceiling. The test bed consists of a net and tether assembly (see in Fig. \ref{fig:Net-test}), a model of debris (a helium blimp, see in Fig. \ref{fig:helium blimp}), a frame that holds the net in an expandable state, a mechanism that raises or lowers the net and tether assembly, and a release mechanism to drop the weights attached to the corners of the net. For simulating debris capture mechanisms, the net is initially kept in an expanded state. The deployment of the net is actuated by the release of corner weights, which fall due to gravity and enclose the target debris \cite{sharf2017experiments}.

\begin{figure}[h]
    \centering
    \includegraphics[width=0.5\linewidth]{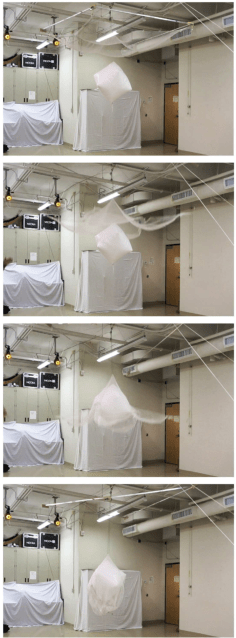}
    \caption{The experiment showing the capture of debris using tether-actuated net closure system. "Reprinted from Acta Astronautica, Vol 139, Inna Sharf and Benjamin Thomsen and Eleonora M. Botta and Arun K. Misra, Experiments and simulation of a net closing mechanism for tether-net capture of space debris, 332-343, Copyright (2017), with permission from Elsevier"  \cite{sharf2017experiments}.}
    \label{fig:net closure with debris}
\end{figure}
The experimental demonstration shows the applicability of a net closure system which is a promising concept for the space debris removal system. The experimental test bed designed in this study is useful for the validation of the dynamics of the net capture system, specifically, the deployment of a net \cite{sharf2017experiments}.

\begin{figure}[h]
    \centering
    \includegraphics[width=0.9\linewidth]{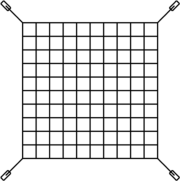}
    \caption{ Schematic of net with mass attached at the corners. "Reprinted from Acta Astronautica, Vol 129, Wojciech Golebiowski and Rafal Michalczyk and Michal Dyrek and Umberto Battista and Kjetil Wormnes, Validated simulator for space debris removal with nets and other flexible tethers applications, 229-240, Copyright (2016), with permission from Elsevier" \cite{GOLEBIOWSKI2016}.}
    \label{fig:experimental net}
\end{figure}

\begin{figure}[h]
    \centering
    \includegraphics[width=0.9\linewidth]{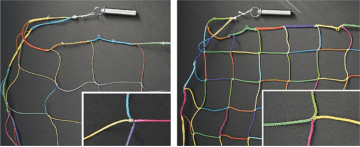}
    \caption{ Assembly technique of knotted and knot-less nets (left and right). "Reprinted from Acta Astronautica, Vol 129, Wojciech Golebiowski and Rafal Michalczyk and Michal Dyrek and Umberto Battista and Kjetil Wormnes, Validated simulator for space debris removal with nets and other flexible tethers applications, 229-240, Copyright (2016), with permission from Elsevier" \cite{GOLEBIOWSKI2016}.}
    \label{fig:Knotted net}
\end{figure}

\begin{figure}[h]
    \centering
    \includegraphics[width=0.9\linewidth]{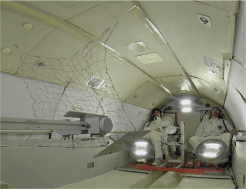}
    \caption{ Snapshot of the camera recording the experiment with
    a thicker 16×16 net (2$^\text{nd}$ case) in the zero gravity phase.). "Reprinted from Acta Astronautica, Vol 129, Wojciech Golebiowski and Rafal Michalczyk and Michal Dyrek and Umberto Battista and Kjetil Wormnes, Validated simulator for space debris removal with nets and other flexible tethers applications, 229-240, Copyright (2016), with permission from Elsevier" 
    \cite{GOLEBIOWSKI2016}.}
    \label{fig:experiment in flight}
\end{figure}

As part of the Clean Space Initiative, another experiment was conducted by SKA Polska under an ESA contract. The experiment used a tether-net capturing system to validate the debris capture mechanism of the scaled-down model of EVISAT. The experiment was conducted in a microgravity environment inside the Falcon-20 aircraft. The trajectory of the deployment of the net and the wrapping of the modeled debris was captured using two fast stereographic camera sets (in total 4 cameras). The setup of the net experiment is shown in Fig. \ref{fig:experimental net}. In this experiment, two cases were tested. In the first case, net with dimensions of 1×1 meter, 10 cm mesh size, 1 mm nylon thread, knotted assembly, spacing: 1.5 meter, and two different velocities of 1.5 m/s and 2.5 m/s ( Fig. \ref{fig:Knotted net}) was considered. In the second case, a net with 1 x 1 meter, 6 cm mesh size, 2 mm nylon thread, knotless assembly, 1.5 meter distance, and two different velocities of 1.5 m/s and 2.5 m/s was considered. (Fig. \ref{fig:Knotted net})\cite{GOLEBIOWSKI2016}.
The experimental setup of thicker net 16x16 (2$^\text{nd}$ test case) was shown in Fig. \ref{fig:experiment in flight}. During the experiment, 22 parabolic flights were carried out by Falcon 20 aircraft. Out of 22, in 20 parabolic flights, there was a successful capture of target that includes the following stages, the deployment and flowing of net, net hitting and wrapping target, and recording of the trajectory\cite{GOLEBIOWSKI2016}.
The experimental results showed that the performance of thicker 16×16 net (2$^\text{nd}$ test case) was worse than thin 10×10 net (1$^\text{st}$ case). The experimental results also validated the dynamics of the net capturing mechanism obtained from the simulation \cite{GOLEBIOWSKI2016}.

\section{\large Preventive Steps}

\newacronym{odmsp}{ODMSP}{Orbital Debris Mitigation Standard Practices}
\newacronym{faa}{FAA}{The Federal Aviation Administration}
\newacronym{noaa}{NOAA}{The National Oceanic and Atmospheric Administration}
\newacronym{iadc}{IADC}{The Inter-Agency Space Debris Coordination Committee}
\newacronym{fcc}{FCC}{Federal Communications Commission}
In year 2002, the Inter-Agency Space Debris Coordination Committee (IADC) developed  "space debris mitigation guidelines" in order to prevent the creation of space debris. These guidelines suggest preventive strategies to control the increased traffic in the popular space orbitals, such as low Earth orbit (LEO), and preserve geostationary orbit from unnecessary space congestion. Thus, allowing sustainable-use of orbitals for future scientific and commercial endeavors. On December 22, 2007, the United Nations passed a resolution 62/217 that endorsed the Space Debris Mitigation Guidelines of the Committee on the Peaceful Uses of Outer Space. The technical content and set of mitigation guidelines are developed by the Inter-Agency Space Debris Coordination Committee (IADC). These guidelines are stated below \cite{UN_space2010}:
\textit{
\begin{itemize}
  \item[] \textbf{Guideline 1}: Limit debris released during normal operations.
  \item[] \textbf{Guideline 2}: Minimize the potential for break-ups during operational phases.
  \item[] \textbf{Guideline 3}: Limit the probability of accidental collision in orbit.
  \item[] \textbf{Guideline 4}: Avoid intentional destruction and other harmful activities.
  \item[] \textbf{Guideline 5}: Minimize potential for post-mission break-ups resulting from stored energy.
  \item[] \textbf{Guideline 6}: Limit the long-term presence of spacecraft and launch vehicle orbital stages in the low-Earth orbit (LEO) region after the end of their mission.
  \item[] \textbf{Guideline 7}: Limit the long-term interference of spacecraft and launch vehicle orbital stages with the geosynchronous Earth orbit (GEO) region after the end of their mission.
\end{itemize}
}
These are voluntary guidelines for the planning and operation of newly designed space missions, as well as for existing space missions wherever possible. The guidelines are not mandatory by international law.

In 2001, the U.S. government created the Orbital Debris Mitigation Standard Practice (\acrshort{odmsp}) to address the increase in space debris in the Earth's orbits. The objective of \acrshort{odmsp} is to prevent the creation of new and long-term space debris by controlling debris released during operations, minimizing the creation of new debris from accidental collisions or explosions, and selecting a safe flight profile and disposing of satellites after their mission\cite{ODMSP}. It also states that spacecraft and rocket bodies must be designed in such a way to minimize the risk of debris creation during the operational phase \cite{ODMSP,migaud2020protecting}.

The Federal Aviation Administration (\acrshort{faa}) is also working in the direction of reducing space debris and has developed regulations for the U.S. launch industry\cite{faa}. According to Title 14 of the Code of Federal Regulations (CFR), Part 415.39, "Safety at the End of Launch," the FAA requires a safety approval for any launch of a launch vehicle. The requirements stated (in 14 CFR part 417.129) below should be fulfilled by the launch vehicle with stages or components that will reach Earth orbit :

\begin{itemize}
    \item After payload separation, there is no unplanned physical contact between the payload and the vehicle or any of its components;
    \item Energy sources such as chemical, pressure and kinetic energy, do not result in the fragmentation of the vehicle or its components, creating new debris.
    \item Stored energy is removed by depleting residual fuel, leaving all fuel line valves open, venting any pressurized system, and leaving all batteries in a permanent discharge state.
\end{itemize}

In June 2004, the Federal Communications Commission (FCC) of the United States (U.S.) issued a comprehensive set of regulations for orbital debris mitigation. These regulations apply to both the licensing of non-US satellites used to offer service in the US as well as the licensing of commercial US satellites. The regulations require that all debris mitigation strategies, including end-of-life measures, be disclosed prior to authorization\cite{kensinger2005united}. The disclosure and operating requirements suggested by the FCC are divided into three categories\cite{kensinger2005united}. 
\begin{itemize}
    \item the disclosures concerning avoiding collisions with other large objects while carrying out routine operations;
    \item post-mission disposal disclosure; and
    \item the disclosure concerning debris control during routine operations; explosion prevention measures; and spacecraft shielding to avoid loss of control due to collisions with small debris.
\end{itemize}   

\newacronym{iaa}{IAA}{International Academy of Astronautics}

Several studies were conducted by the International Academy of Astronautics (\acrshort{iaa}) on the current status and stability of space debris to address the possibilities of space traffic management and subsequently work on the mitigation of the creation of space debris\cite{klinkrad2009space}.
\newacronym{gto}{GTO}{geostationary transfer orbit}
\newacronym{sso}{SSO}{super-synchronous orbit}
According to the 2001 IAA \textit{Position Paper on Space Debris}, the management phases for reducing and eliminating space debris are divided into three categories. The first category requires immediate actions, and these actions are \cite{Flury2001}: 
\begin{itemize}
    \item [1.]There will be no deliberate spacecraft breakups, resulting in debris in orbit.
    \item [2.]Efficient design and development of spacecrafts and satellites to reduce mission-related debris.
    \item [3.]The removal of internal energy from the upper stage and spacecraft at the end of a mission in any Earth orbit.
    \item [4.] Optimization of Geostationary transfer orbit (GTO) parameter to reduce upper stage lifetime.
    \item [5.] Transferring decommissioned geostationary satellites to a disposal orbit.
    \item[6.]  If separation of ABM (Apogee-Boost Motor) is needed from a geostationary spacecraft, it should be done in super-synchronous orbit.
    \item [7.] Passivation and disposal of upper stages used to move geostationary satellites, should be done in orbit atleast 300km above the geostationary orbit.
\end{itemize}
The second category requires a change in operational procedures or hardware and is intended to remove abandoned rockets and spacecraft from LEO. This results in a significant reduction in space debris. The actions needed in second category are:
\begin{itemize}
    \item [1.] Limiting the orbital lifetime of upper stages and defunct satellites to 25 years in LEO.
    \item [2.] Transfer of upper stages and defunct satellite to disposal orbit after completion of mission.
\end{itemize}

Debris management options in the third category
generally require new development and demonstration of
their suitability (technical feasibility,
cost-effectiveness). Their implementation is the priority with five phases designated as follows\cite{Flury2001, bockstiegel1990space,bonnal2014requirements}: 
\begin{itemize}
    \item [1.]Development of the capability for deorbit with propulsion.
    \item [2.] Drag augmentation systems to increase drag for deorbiting.
    \item [3.] Development of grappling, tumbling, and tethering devices for removal.
    \item [4.] Development of laser removal.
    \item [5.] Development of effective sweepers that can avoid collisions.
\end{itemize}

The above guidelines or suggestions made by various space agencies and regulating bodies are helpful in preventing the creation of new debris.

\section{\large Discussion and Conclusion}

The threat posed by today's space activities stems from the past when it was believed that "Space" offers unlimited opportunities for human activities \cite{abdollahi2016international}. Now, because of human interference, this issue has become a critical problem that threatens the space environment. As we know, there are only a few countries that are capable of launching rockets or satellites into space. However, the number of objects in space has increased dramatically over the years, due to numerous space activities, including satellite launches and the conduction of anti-satellite missile tests. Therefore, there is a serious concern about the removal of objects (debris) from Earth's orbit, especially from the low earth orbit (LEO). Several researchers and space agencies are constantly working to design and develop efficient systems for the removal of space debris. In this article, we explore the current status of space debris in the Earth's orbits by analysing the space object databases provided by the European Space Agency (ESA) and NASA. We also study the characteristics of space debris (such as number, size, speed, mass, position, and cross-sectional area), that would be helpful for the development of debris removal systems.  Our study suggests that the speed of the debris in Earth's orbit is more important than its mass because the high kinetic energy of the debris can cause considerable damage. It is also found that de-orbiting defunct satellites to a higher orbit can reduce the likelihood of collisions.
 In the present article, we highlight the mitigation strategies to constrain the growth of space debris and discuss the recent activities of space agencies for the removal of space debris, including NASA's "Clear Space-01" and ESA's "Elsa-D satellite" projects. We share information related to various debris-capturing methods and found that net-based debris-capturing is an efficient and cost-effective method. To support the development of such systems, we present the current progress in modelling activities of net-based debris-capturing methods. Finally, we present preventive measures provided by international space agencies, including the Inter-Agency Space Debris Coordination Committee (IADC) for controlling increased traffic in popular space orbitals.
 
 \section*{Acknowledgement}
 The author (M.B.) is thankful to ..... Unisalento for financial support. The author (R.S.) thanks Konstantinios Pagkalis for fruitful discussion.






\printglossary[type=\acronymtype]
\bibliographystyle{unsrtnat}
\bibliography{name}
\clearpage

\end{document}